\documentclass{BZS}
\usepackage[super,comma,sort&compress]{natbib}
\usepackage[T1]{fontenc}
\usepackage[flushleft]{threeparttable}
\usepackage{hyperref}
\hypersetup{colorlinks,citecolor=blue}
\usepackage{mathptmx}
\usepackage{multirow}
\usepackage{amsmath}

\newcommand{\sous}[1]{\ensuremath{_{\textrm{#1}}}}

\begin{document}
\setlength{\parskip}{1em}
\pagestyle{plain}

\justifying
\title{Ruddlesden-Popper chalcogenides push the limit of mechanical stiffness and glass-like thermal conductivity in single crystals}

\maketitle


\author{Md Shafkat Bin Hoque, Eric R. Hoglund, Boyang Zhao, De-Liang Bao, Hao Zhou, Sandip Thakur, Eric Osei-Agyemang, Khalid Hattar, Ethan A. Scott, Mythili Surendran, John A. Tomko, John T. Gaskins, Kiumars Aryana, Sara Makarem, Adie Alwen, Andrea Hodge, Ganesh Balasubramanian, Ashutosh Giri, Tianli Feng, Jordan A. Hachtel, Jayakanth Ravichandran, Sokrates T. Pantelides, and Patrick E. Hopkins}


\dedication{}

\begin{affiliations}
\normalsize Md Shafkat Bin Hoque, John A. Tomko, Kiumars Aryana\\
Department of Mechanical and Aerospace Engineering, University of Virginia, Charlottesville, Virginia 22904, USA\hfill

\normalsize Eric R. Hoglund\\
Department of Materials Science and Engineering, University of Virginia, Charlottesville, Virginia 22904, USA\\
Center for Nanophase Materials Sciences, Oak Ridge National Laboratory, Oak Ridge, Tennessee 37830, USA\hfill

\normalsize Boyang Zhao, Mythili Surendran, Adie Alwen, Andrea Hodge\\
Mork Family Department of Chemical Engineering and Materials Science, University of Southern California, Los Angeles, California 90089, USA\hfill

\normalsize De-Liang Bao\\
Department of Physics and Astronomy, Vanderbilt University, Nashville, Tennessee 37235, USA\hfill

\normalsize Hao Zhou, Tianli Feng\\
Department of Mechanical Engineering, University of Utah, Salt Lake City, Utah 84112, USA\hfill

\normalsize Sandip Thakur, Ashutosh Giri\\
Department of Mechanical, Industrial, and Systems Engineering, University of Rhode Island, Kingston, Rhode Island 02881, USA\hfill

\normalsize Eric Osei-Agyemang\\
Department of Materials Design and Innovation, University at Buffalo, The State University of New York, Buffalo, New York 14260, USA\hfill

\normalsize Ganesh Balasubramanian\\
Department of Mechanical Engineering and Mechanics, Lehigh University, 19 Memorial Drive West, Bethlehem, Pennsylvania 18015, USA\hfill

\normalsize Sara Makarem\\
Department of Materials Science and Engineering, University of Virginia, Charlottesville, Virginia 22904, USA\hfill

\normalsize Khalid Hattar\\
Nuclear Engineering, University of Tennessee, Knoxville, Tennessee 37996, USA\\
Sandia National Laboratories, Albuquerque, New Mexico 87185, USA\hfill

\normalsize Ethan A. Scott\\
Sandia National Laboratories, Albuquerque, New Mexico 87185, USA\hfill

\normalsize John T. Gaskins\\
Laser thermal analysis, Charlottesville, Virginia 22902, USA\hfill

\newpage
\normalsize Jordan A. Hachtel\\
Center for Nanophase Materials Sciences, Oak Ridge National Laboratory, Oak Ridge, Tennessee 37830, USA\hfill

\normalsize Jayakanth Ravichandran\\
Mork Family Department of Chemical Engineering and Materials Science, University of Southern California, Los Angeles, California 90089, USA\\
Email: j.ravichandran@usc.edu

\normalsize Sokrates T. Pantelides\\
Department of Physics and Astronomy, Vanderbilt University, Nashville, Tennessee 37235, USA\\
Department of Electrical and Computer Engineering, Vanderbilt University, Nashville, Tennessee 37235, USA\\
Email: pantelides@vanderbilt.edu

\normalsize Patrick E. Hopkins\\
Department of Mechanical and Aerospace Engineering, University of Virginia, Charlottesville, Virginia 22904, USA\\
Department of Materials Science and Engineering, University of Virginia, Charlottesville, Virginia 22904, USA\\
Department of Physics, University of Virginia, Charlottesville, Virginia 22904, USA\\
Email: phopkins@virginia.edu

Md Shafkat Bin Hoque, Eric R. Hoglund, and Boyang Zhao contributed equally to this work.

Jayakanth Ravichandran, Sokrates T. Pantelides, and  Patrick E. Hopkins are corresponding authors. 

\end{affiliations}

\newpage
\textbf{\Large Abstract}
\begin{abstract}

Insulating materials featuring ultralow thermal conductivity for diverse applications also require robust mechanical properties. Conventional thinking, however, which correlates strong bonding with high atomic-vibration-mediated heat conduction, led to diverse weakly bonded materials that feature ultralow thermal conductivity and low elastic moduli. One must, therefore, search for strongly-bonded single crystals in which heat transport is impeded by other means. Here, we report intrinsic, glass-like, ultralow thermal conductivity and ultrahigh elastic-modulus/thermal-conductivity ratio in single-crystalline Ruddlesden-Popper Ba$_{n+1}$Zr$_n$S$_{3n+1}$, n = 2,3, which are derivatives of BaZrS$_{3}$. Their key features are strong anharmonicity and intra-unit-cell rock-salt blocks. The latter produce strongly bonded intrinsic superlattices, impeding heat conduction by broadband reduction of phonon velocities and mean free paths and concomitant strong phonon localization. The present study initiates a paradigm of “mechanically stiff phonon glasses”.

\end{abstract}



\section{Introduction}

Designing materials with ultralow thermal conductivity ($\kappa$) without reducing their density and degrading their mechanical properties typically evades century-old theories on microscopic heat conduction. In non-metallic crystals, the primary modes of thermal transport are lattice vibrations, namely phonons. Dating back to theories originally pioneered by Peierls, Leibfreid and Schlomann, Einstein, and Debye, the thermal transport of atomic vibrations are directly related to the interatomic bond strengths.\cite{kittel1949interpretation,ioffe1960non,slack1979thermal,cahill1992lower,clarke2003materials,agne2018minimum,beekman2017inorganic,mukhopadhyay2018two,kim2021pushing} Hence, "ultralow" thermal conductivity in solids often comes at the cost of “weak” interatomic-bond strength and “soft” elastic modulus (\textit{E}), limiting their mechanical performance.\cite{chiritescu2007ultralow,duda2013exceptionally,wang2013ultralow,braun2018charge,qian2021phonon} In recent years, a family of halide perovskite-structure crystalline solids has been found to feature ultralow thermal conductivities, of order 0.2-0.3 W m$^{-1}$ K$^{-1}$, but their weak van der Waals (vdW) bonding leads to very low values of elastic and shear moduli.\cite{elbaz2017phonon,acharyya2020intrinsically,acharyya2022glassy,acharyya2023extended,christodoulides2021signatures} Breaking this Pareto-normality in the design of crystals to create ultrahigh \textit{E}/$\kappa$ materials must, therefore, involve different strategies than those traditionally proposed to limit phonon transport. Such decoupling would have major implications in many technological applications, such as thermal barrier coatings,\cite{yang2019diffused} thermoelectrics,\cite{snyder2011complex,ravichandran2014crossover} and memory devices.\cite{aryana2021tuning,gibson2021low} 

To overcome the barriers of achieving ultrahigh \textit{E}/$\kappa$ materials and create new directions for achieving single crystals with ultralow thermal conductivity, one seeks to inhibit the propagation of phonons in a “hard” lattice. In perfect crystals, phonons travel with intrinsic group velocities and are inhibited by phonon-phonon scattering caused by anharmonic effects. Historically, chalcogenides (S, Se, and Te) have been investigated for ultralow theral conductivity because of strong anharmonicity (e.g., PbSe, Bi$_2$Te$_3$, SnSe).\cite{chiritescu2009lower,manley2019intrinsic,ong2017orientational,zhao2014ultralow,su2022high} However, many of these materials tend to crystallize in layered structures with vdW interactions, leading to poor mechanical properties across the layers due to ease of shearing. Thus, covalently-bonded layered chalcogenides present a potential pathway to low thermal conductivity \textit{and} high elastic modulus in the direction across the layers. We, therefore, present Ruddlesden-Popper (RP) sulfides\cite{niu2018optimal,niu2019crystal} as potential candidates for this scenario. Unlike some RP halide perovskites that are bonded by weak vdW forces,\cite{christodoulides2021signatures,acharyya2020intrinsically} the RP sulfides hold both the promise of high anharmonicity and strong bonding in a layered, superlattice-like structure with intrinsic interfaces.  

Herein, we report on the ultralow thermal conductivity of RP phases Ba$_{n+1}$Zr$_n$S$_{3n+1}$, n = 2 and 3 of barium zirconium sulfide (BaZrS$_{3}$)\cite{niu2019crystal,li2019band} single crystals, enabled by strong anharmonicity and a large fraction of localized and low-velocity vibrational modes throughout the entire vibrational spectrum, thus achieving broadband attenuation of thermal transport. We reveal the origin of the ultralow thermal conductivities in these RP derivatives of the chalcogenide perovskite BaZrS$_{3}$ using a combination of experiments and first-principles- and machine learning-driven computational approaches. Unlike previously studied vdW layered crystals, including the RP halides,\cite{christodoulides2021signatures,acharyya2020intrinsically}  the strong Ba-S and Zr-S chalcogenide bonds across the rock-salt layers and the (BaZrS$_{3}$)$_n$ layers bring elastic-moduli values of the RP phases nearly an order of magnitude higher than other ultralow-thermal-conductivity inorganic crystals. As a result, the RP phases of these single crystals exhibit record setting values of \textit{E}/$\kappa$ for any single-crystalline material discovered to date, while maintaining glass-like, ultralow thermal conductivities. Such a useful combination of thermal and mechanical properties makes the materials highly important in the fields of thermal barrier coatings and thermoelectric materials.\cite{toberer2012advances}

\section{Results and discussion}

\begin{figure}[hbt!]
\begin{center}
\includegraphics[scale = 1]{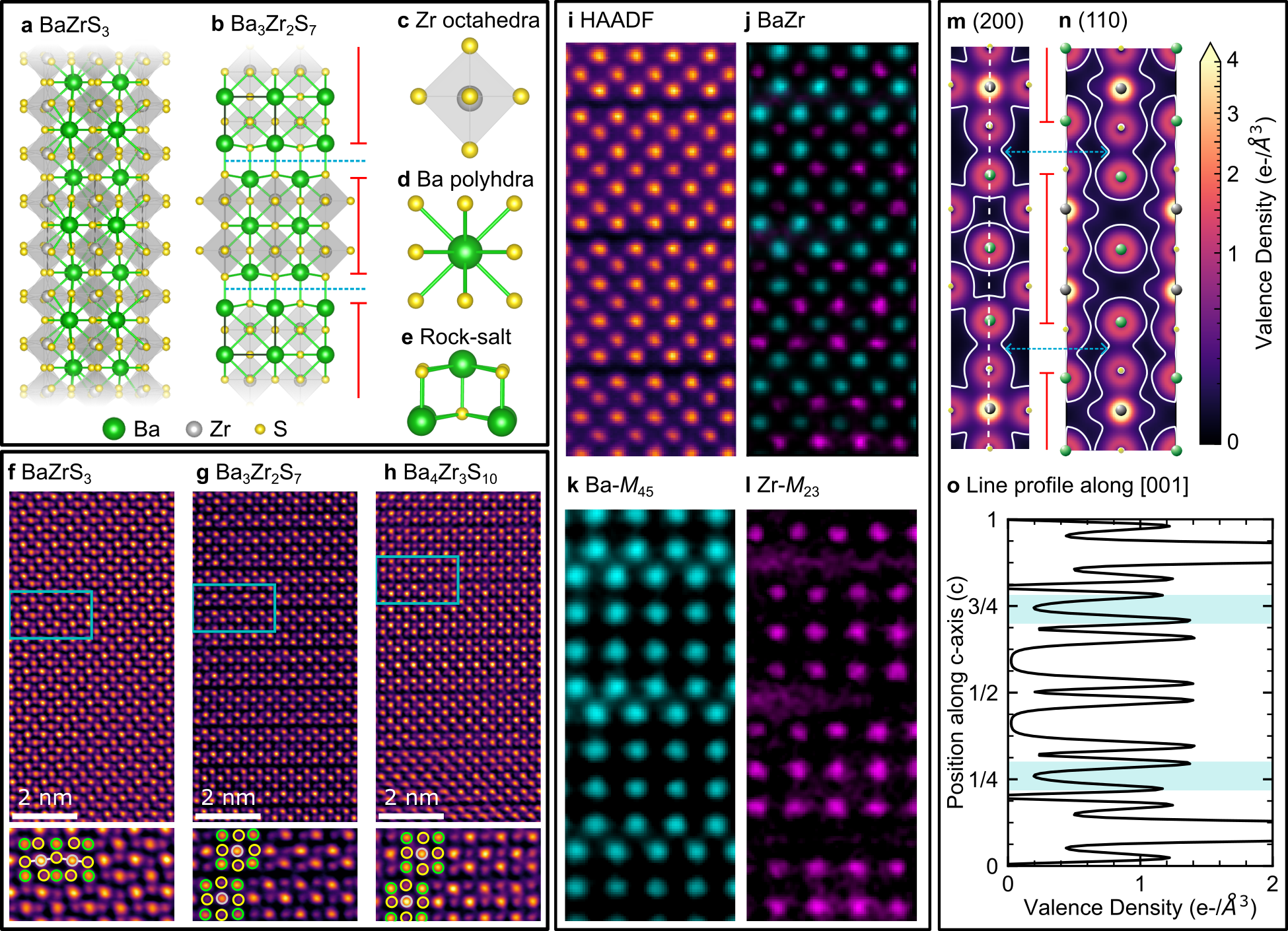}
\caption{Structure of perovskite BaZrS$_{3}$ and Ruddlesden-Popper Ba$_{n+1}$Zr$_n$S$_{3n+1}$. Ball-and-stick model of (a) \textit{Pnma} perovskite BaZrS$_{3}$ and (b) \textit{I4mmm} Ruddlesden-Popper Ba$_{3}$Zr$_{2}$S$_{7}$ showing grey ZrS$_{6}$ octahedra and green BaS bonds. Red markers in (b) indicate the perovskite blocks of the RP phase separated by the rock-salt blocks. Blue dotted lines between BaS atomic planes indicate the midplane of the rock-salt blocks. Ball-and-stick model of a (c) Zr octahedra, (d) undistorted Ba polyhedral, and (e) rock-salt building block resulting from layering in Ba$_{3}$Zr$_{2}$S$_{7}$. Integrated differential phase contrast images of the (f) BaZrS$_{3}$, (g) Ba$_{3}$Zr$_{2}$S$_{7}$, (h) Ba$_{4}$Zr$_{3}$S$_{10}$ crystals. Enlarged regions from the cyan annotations are shown below each image. In the enlargements, two perovskite unit cells are annotated with Ba (green), Zr (grey) and S (yellow) circles. (i) Z-contrast image and (j) BaZr composite image from a STEM-EELS spectrum image. Intensity maps of (k) Ba-M$_{45}$, (l) Zr-M$_{23}$ background-subtracted edges. (m,n) Section views along (200) and (110) of the valence electron density of the RP-phase Ba$_{3}$Zr$_{2}$S$_{7}$ calculated by DFT, respectively. Red marks and blue dashed lines help to correlate atomic structure to that of panel (b). (o) A line profile of the valence electron density along the white dashed line in (m). Blue bars illustrate the rock-salt-block regions.}
\label{Figure 1}
\end{center}
\end{figure}

To investigate the impact of sub-unit cell structures on thermal conductivity, we consider perovskite BaZrS$_{3}$ and two RP phases, Ba$_{3}$Zr$_{2}$S$_{7}$ and Ba$_{4}$Zr$_{3}$S$_{10}$. The perovskite shown in Figure 1(a) consists of tilted ZrS$_{6}$ octahedra (Figure 1(c)) and BaS$_{12}$ polyhedra (Figure 1(d)). The RP-phase Ba$_{3}$Zr$_{2}$S$_{7}$ shown in Figure 1(b) contains two perovskite sections (red brackets) that are separated by rock-salt packed layers (Figure 1(e)). Ba$_{4}$Zr$_{3}$S$_{10}$ differs from Ba$_{3}$Zr$_{2}$S$_{7}$ by adding one more BaS and ZrS$_{2}$ atomic layer to each perovskite section. Figure 1(f) shows an iDPC image of BaZrS$_{3}$ and Figures 1(g,h) show the layered periodic stacking of perovskite layers in the two RP phases with the enlarged regions of interest, emphasizing one of two rock-salt layers that are present in a single unit cell. The elemental maps generated by electron-energy-loss spectroscopy (EELS) and the corresponding atomic-number-contrast (Z-contrast) image shown in Figure 1(i-l) show the high degree of chemical ordering in each sublattice and the change in local symmetry at the rock-salt layers.

To understand the bonding in the RP phases, density functional theory (DFT) calculations were performed on Ba$_{3}$Zr$_{2}$S$_{7}$ to calculate the valence electron density, as shown in Figure 1(m,n). A line profile along the dashed white line in Figure 1(m) is shown in Figure 1(o) for a more quantitative evaluation. The valence electron density within the rock-salt regions is non-zero and comparable to those inside the perovskite blocks, which suggests similar intra- and inter-perovskite-block bonding strength. In other words, the bonding in the rock-salt regions, namely across the “gaps” highlighted in Figure 1(b), is not of the weak, vdW type. The overall strong bonding is also reflected in the calculated elastic moduli, which have comparable values along the cross-plane and in-plane directions. The elastic-moduli values are three to four times higher than those observed in other ultralow-thermal-conductivity halide perovskites, such as, Cs$_3$Bi$_2$I$_{6}$Cl$_{3}$,\cite{acharyya2022glassy} Cs$_3$Bi$_2$I$_{9}$,\cite{acharyya2023extended} the RP-phase Cs$_2$PbI$_2$Cl$_{2}$.\cite{acharyya2020intrinsically} and several metal halide perovskites (see Supplemental Table S3). The presence of strong intra- and inter-perovskite-block bonding strength in the sulfide RP phases described here mitigate the role of bond strength in the observed ultralow thermal conductivity.

\begin{figure}[hbt!]
\begin{center}
\includegraphics[scale =0.28]{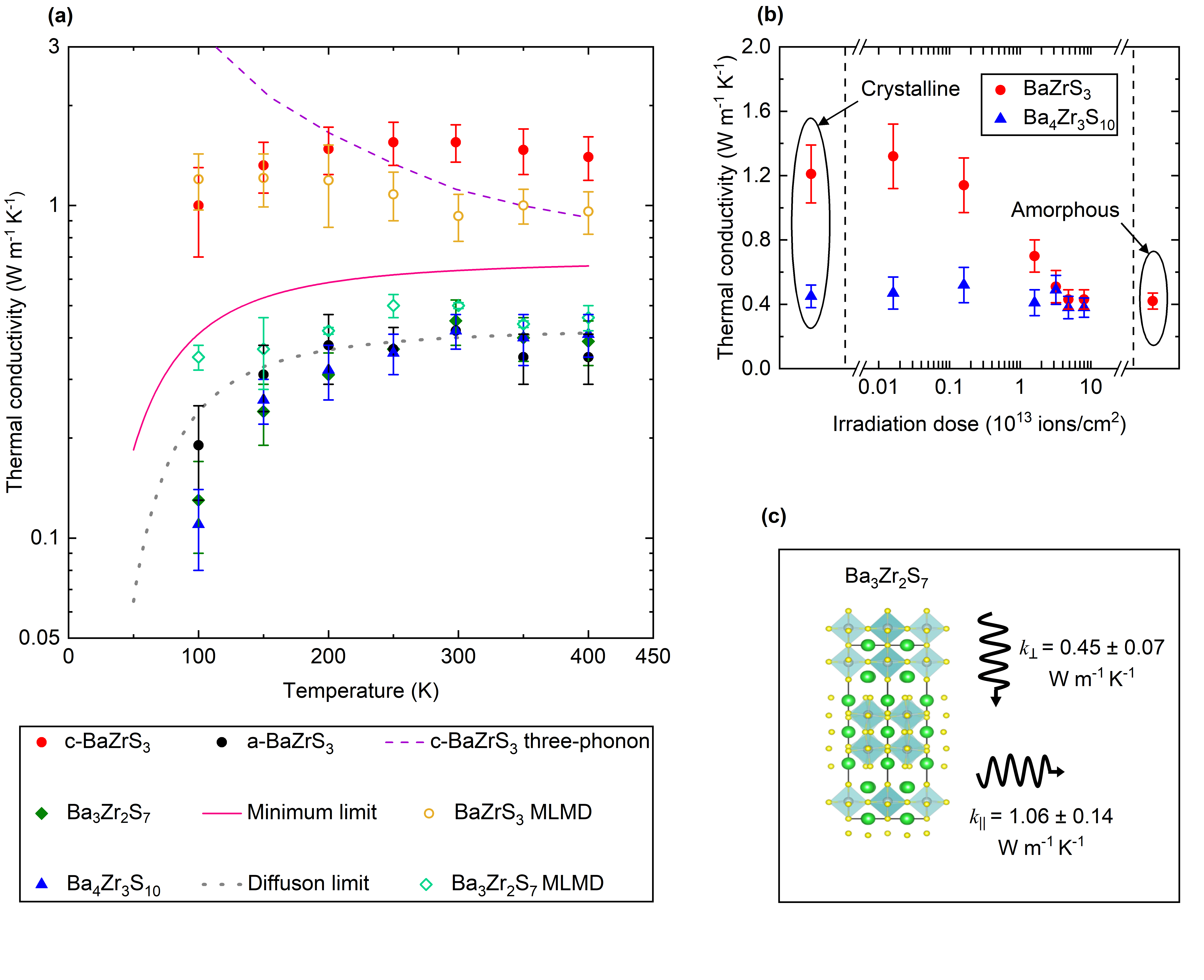}\\
\caption{Thermal conductivities – experimental and simulated. (a) TDTR-measured cross-plane thermal conductivity of BaZrS$_{3}$ and its RP derivatives. c-BaZrS$_{3}$ and a-BaZrS$_{3}$ denote crystalline and amorphous BaZrS$_{3}$, respectively. For comparison, we also include the three-phonon and MLMD predicted thermal conductivity of BaZrS$_{3}$ and RP phases. The three-phonon prediction is adopted from Osei-Agyemang \textit{et al.}\cite{osei2019understanding} The minimum limit and diffuson limit refer to Cahill \textit{et al.}\cite{cahill1992lower} and Agne \textit{et al.}’s\cite{agne2018minimum} theoretical models. (b) Thermal conductivity distribution of BaZrS$_{3}$ and RP phases as a function of heavy ion-irradiation doses. The thermal conductivity of the crystalline BaZrS$_{3}$ is lower compared to panel (a) due to the presence of nano-domains (see Supporting Information for details). (c) Anisotropic thermal conductivity of Ba$_{3}$Zr$_{2}$S$_{7}$ measured by TDTR.}   
\label{Figure 2}
\end{center}
\end{figure}

To understand phonon transport through the structures, we measured the cross-plane (\textit{c}-axis) thermal conductivity using time-domain thermoreflectance (TDTR) from 100 to 400 K, as shown in Figure 2(a). The thermal conductivities exhibit several unusual features for single-crystalline materials. First, the thermal conductivity of crystalline BaZrS$_{3}$ increases from 100 to 250 K, then remains relatively temperature independent. Such trend is observed in amorphous materials and disordered crystals, but it is rare in single crystals.\cite{beekman2017inorganic} The weak or negligible temperature dependence cannot be explained by prior first-principles three-phonon scattering calculations\cite{osei2019understanding} (the purple dashed curve in Figure 2a). In contrast, our DFT-based machine-learning-interatomic-potential (MLIP)-driven molecular-dynamics (MD) simulation results (orange open dots in Figure 2a) show good overall agreement with experimentally measured data, with only small differences above 200 K. This agreement demonstrates the accuracy of MLIP-MD (or MLMD) simulations, which implicitly capture all the atomic-vibration contributions to thermal conductivity, including the effects of three-phonon scattering, four-phonon scattering, finite-temperature phonon renormalization, temperature correction to force constant, and diffuson contributions.\cite{tiwari2024accurate,kulnitskiy2024effect} This agreement also confirms that the glass-like thermal conductivity trend of single-crystalline BaZrS$_{3}$ is not from extrinsic defects, but rather mechanisms intrinsic to the crystal. However, the exact mechanism behind the temperature trend in the thermal conductivity of single-crystalline BaZrS$_{3}$ remains an open question. Supporting Figure S9 shows that perovskites have been reported to exhibit both glass- and crystalline-like thermal conductivity trends. Isolating the mechanisms that underlie the thermal conductivities observed in this class of materials is beyond the scope of the present study, as we focus on the RP phases.

The RP phases Ba$_{3}$Zr$_{2}$S$_{7}$ and Ba$_{4}$Zr$_{3}$S$_{10}$ possess ultralow thermal conductivities, i.e., 0.45 ± 0.07 and 0.42 ± 0.05  W m$^{-1}$ K$^{-1}$, respectively, over a relatively large temperature range. These values are $\sim$3.5 times lower than those of single-crystalline BaZrS$_{3}$, in agreement with the Agne \textit{et al.}’s diffuson limit,\cite{agne2018minimum} and lower than Cahill \textit{et al.}’s glass limit.\cite{cahill1992lower} These two limits are two of the most commonly used theoretical models to predict the lowest possible thermal conductivity of a crystalline material. Additionally, the thermal conductivities of Ba$_{3}$Zr$_{2}$S$_{7}$ (green diamonds) and Ba$_{4}$Zr$_{3}$S$_{10}$ (blue triangles) show glass-like temperature trends, comparable to that of amorphous BaZrS$_{3}$ (a-BaZrS$_{3}$) (black circles). The MLMD thermal conductivity simulations of Ba$_{3}$Zr$_{2}$S$_{7}$ show quantitative agreement with the experimental data. The cross-plane thermal conductivities of the sulfide RP phases are even lower than those of vdW layered materials, except when thin film samples and interlayer rotations are involved.\cite{chiritescu2007ultralow, kim2021extremely} On the other hand, the sulfide RP phases feature far superior mechanical properties of all vdW layered materials.  

Defects in materials have also been known to lead to ultralow thermal conductivity and glass-like temperature trends.\cite{beekman2017inorganic} To rule out the influence of defect scattering on our experimental data, we irradiated the BaZrS$_{3}$ and Ba$_{4}$Zr$_{3}$S$_{10}$ crystals with high energy gold ions. The measured cross-plane thermal conductivities of the heavily ion-irradiated crystals as a function of ion dose are shown in Figure 2(b). The thermal conductivity of BaZrS$_{3}$ exhibits a sigmoidal reduction, typically characteristic of irradiated crystalline materials.\cite{scott2020orders,scott2021reductions} At low doses, irradiation introduces low concentrations of clustered point defects and vacancies. The overall crystal structure remains relatively unchanged, whereby the thermal conductivity is nearly constant. At high doses, point-defect concentrations increase and damaged regions overlap, which gradually decreases the thermal conductivity to that of an amorphous solid. These observations, together with the agreement of the simulations of the thermal conductivity of c-BaZrS$_{3}$ with the experimental data, also prove the high quality of our unirradiated BaZrS$_{3}$ single-crystals. Large concentrations of defects in the unirradiated crystals would have caused the thermal conductivity to drop below the shown value. 

Compared to BaZrS$_{3}$, a completely different thermal conductivity trend is observed vs. ion dose in the RP phases; the thermal conductivity of the Ba$_{4}$Zr$_{3}$S$_{10}$ crystals remains nearly constant regardless of gold ion dose. TEM micrographs show that the layering of Ba$_{4}$Zr$_{3}$S$_{10}$ crystals remains uninterrupted throughout the range of doses, although high doses can introduce amorphous pockets (see Supporting Information). In vdW layered materials with interlayer rotation, ion irradiation can lead to increased thermal conductivity due to increases in interatomic bonding.\cite{chiritescu2007ultralow} However, due to the already strong bonding of the RP phases, no such trend is observed in the present study. The resistance of the RP crystals to irradiation damage makes them a highly suitable thermal barrier coating in deep space applications in radiation environments.

To gain insight into the role of anisotropy in ultralow thermal conductivity of the RP phases, we measured the thermal conductivity of  Ba$_{3}$Zr$_{2}$S$_{7}$ along the in-plane direction (perpendicular to \textit{c}-axis) at room temperature as shown in Figure 2(c).  The in-plane thermal conductivity is 1.06 $\pm$ 0.14 W m$^{-1}$ K$^{-1}$ which is $\sim$2.5 times higher than the cross-plane thermal conductivity (0.45 $\pm$ 0.07 W m$^{-1}$ K$^{-1}$). Noteworthy, the MLMD-simulated in-plane and cross-plane thermal conductivities are 1.02 $\pm$ 0.17 and 0.53 $\pm$ 0.02 W m$^{-1}$ K$^{-1}$, respectively, showing excellent  quantitative agreement with the experimental anisotropy. The major structural difference between the in-plane and cross-plane directions is the periodic rock-salt and perovskite layers, relative to continuous layers in-plane. This result suggests that, despite the strong bonding, the rock-salt layers are causing the ultralow and anisotropic thermal conductivity in the RP phases.

\begin{figure}[hbt!]
\begin{center}
\includegraphics[scale = 1]{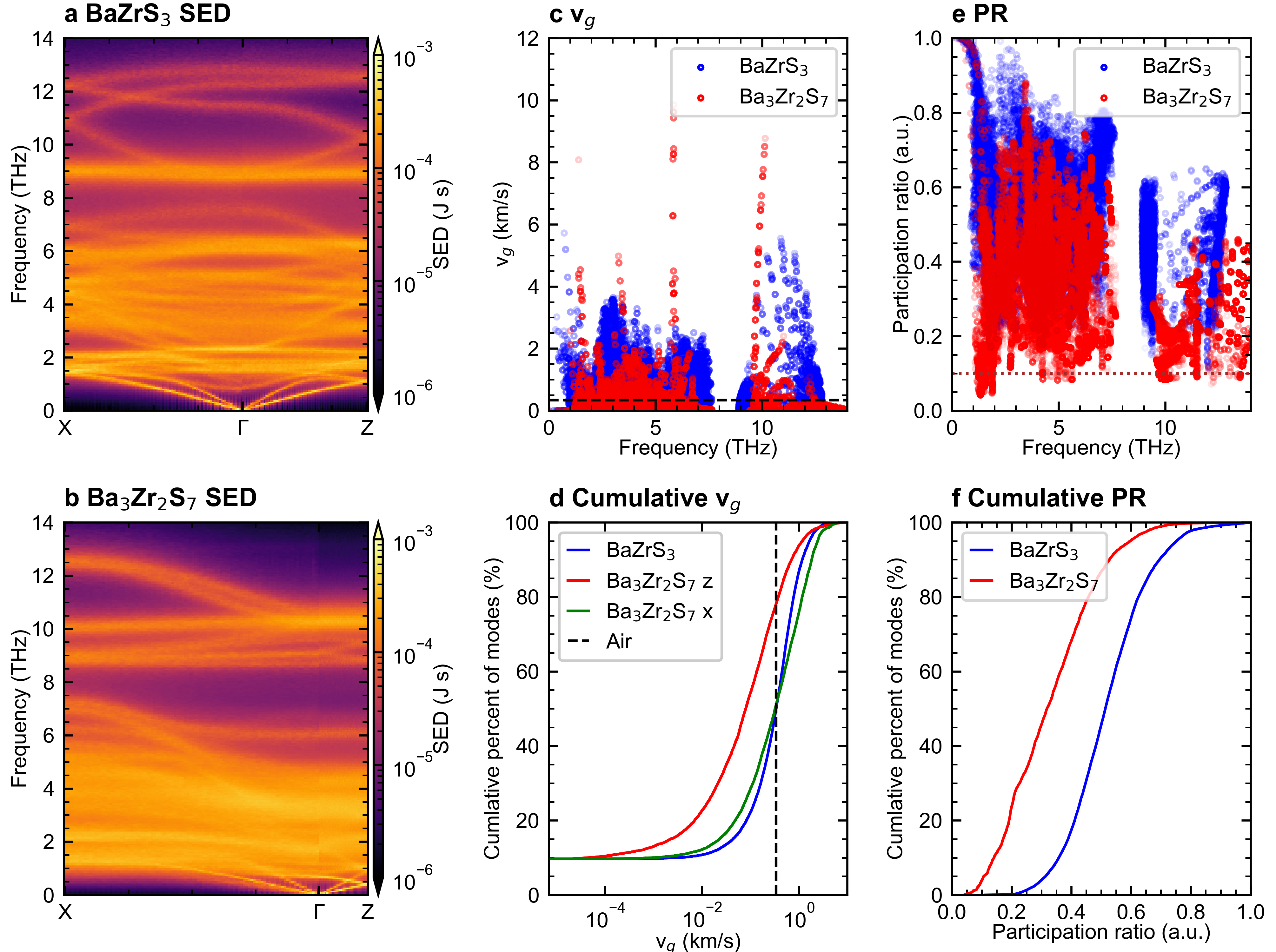}\\
\caption{Evaluation of factors that influence thermal conductivity. Spectral energy density calculated from MLMD for BaZrS$_{3}$ and (b) Ba$_{3}$Zr$_{2}$S$_{7}$. (c) v$_{g}$ of the BaZrS$_{3}$ and RP structure along the cross-plane direction as a function of frequency with the dashed black line serving as a reference for air. (d) Cumulative number of phonon modes having up to a given group velocity as a function of frequency for the crystals along cross- (\textit{z}) and in- (\textit{x}) plane directions. (e) Participation ratio (PR) as a function of frequency for BaZrS$_{3}$ and (b) Ba$_{3}$Zr$_{2}$S$_{7}$. (f) Cumulative number of phonon modes having up to a given PR as a function of frequency.}
\label{Figure 3}
\end{center}
\end{figure}

To understand how the introduction of rock-salt layering leads to the ultralow thermal conductivities in the RP phases compared to that of crystalline BaZrS$_{3}$, we consider three factors that may contribute to ultralow thermal conductivity: 1) anharmonic scattering, 2) decreased phonon group velocities (v$_{g}$), and 3) phonon localization within unit cells.\cite{thomas2010predicting}

To assess the role of anharmonic scattering, we calculated the spectral energy density (SED) of BaZrS$_{3}$ and Ba$_{3}$Zr$_{2}$S$_{7}$ based on MLMD, as shown in Figure 3(a,b). Compared to a typical semiconductor (e.g., silicon, gallium nitride), the SEDs of both BaZrS$_{3}$ and Ba$_{3}$Zr$_{2}$S$_{7}$ show much more blurred and broadened linewidths, indicating strong anharmonicity and large phonon scattering rates.\cite{de2009thermal,thomas2010predicting} These features could be a critical reason for the ultralow thermal conductivity. The large overlapping of branches due to the anharmonic broadening also indicates that the interband tunneling (i.e., diffusons) should be significant, based on the Wigner formalism.\cite{simoncelli2019unified,simoncelli2023thermal} This feature could be a reason for the glass-like thermal conductivity exhibited by the experimental data. Recent calculations using the Wigner formalism show that diffuson contributes 30$\%$ to thermal conductivity of BaZrS$_{3}$. We expect the diffuson contribution to be even larger in Ba$_{3}$Zr$_{2}$S$_{7}$ since the broadening of SED is more significant. 

Group velocity (v$_{g}$) is another important parameter to understand the diffusivity of vibrational modes in a material.\cite{braun2018charge} The group velocities along the cross-plane direction (v$_{g,z}$) of BaZrS$_{3}$ and Ba$_{3}$Zr$_{2}$S$_{7}$ are compared in Figure 3(c). The latter shows ultralow group velocities compared to the former at nearly all frequencies, except for very few optical modes, whose fraction is very small. To quantify the number of modes having ultralow group velocities, we plot the cumulative number of phonon modes as a function of group velocity in Figure 3(d). We find that 80$\%$ of phonon modes of Ba$_{3}$Zr$_{2}$S$_{7}$ have a v$_{g,z}$ lower than the sound speed of air, which is astonishing. In comparison, the \textit{x} component shows a larger group velocity, which is similar to that of BaZrS$_{3}$. This feature would indicate that the in-plane thermal conductivity of Ba$_{3}$Zr$_{2}$S$_{7}$ is similar to that of BaZrS$_{3}$ but the cross-plane thermal conductivity would be much lower, which is consistent with both the experimental observations and the simulations. Thus, the ultralow phonon velocities, induced by the presence of periodic building blocks in unit cells, is another key contributing factor to the ultralow thermal conductivities.

The third possible factor contributing to ultralow thermal conductivity is phonon localization. For that, we calculate the participation ratio (PR) of the vibrational modes, which represents the spatial localization of phonon waves within a unit cell. Localized vibrational modes are usually defined as having a participation ratio lower than 0.1.\cite{seyf2016method} As shown in Figure 3(e,f), the participation ratio of Ba$_{3}$Zr$_{2}$S$_{7}$ is significantly lower than that of BaZrS$_{3}$ across all frequencies. It is noteworthy that some low-frequency modes (<2 THz) in the RP phase have a participation ratio that is smaller than 0.1, which is comparable to the localization expected for locons in amorphous materials.\cite{lv2016non,seyf2016method} This result provides evidence that the presence of rock-salt building block layers in the RP phases causes a significant number of vibrational modes to become highly localized.\cite{acharyya2023extended} 

To further show the localization of the vibrational modes, we estimate the average mean free path of phonons in Ba$_{3}$Zr$_{2}$S$_{7}$ in Supporting Figure S10. Assuming the diffuson thermal conductivity is zero, the average mean free path of phonons is estimated as 1 nm, to match with experimental thermal conductivity. Since diffuson contribution is nonzero, the actual mean free path of phonons should be smaller than 1 nm, the inter-gap thickness. These considerations indicate that the phonons are localized inside the rock-salt layers of Ba$_{3}$Zr$_{2}$S$_{7}$ by the gaps, being consistent with the participation ratio results.

In summary, the presence of intra-unit-cell rock-salt blocks in the sulfide RP phases derived from BaZrS$_{3}$ and the corresponding selenide and telluride RP phases effectively produces strongly bonded, intrinsic superlattices with different periodicities and interfacial regions that largely reduce phonon velocities and mean free paths, inducing strong localization. Combined with the strong anharmonicity that is known to be intrinsic to chalcogenides,\cite{chiritescu2009lower,manley2019intrinsic,ong2017orientational,zhao2014ultralow,su2022high} the chalcogenide RP phases are a class of single-crystalline materials that can achieve broadband restriction of thermal transport, leading to ultralow thermal conductivities, while sustaining high elastic moduli and hence high \textit{E}/$\kappa$ ratios.

In Figure 4, we compare the \textit{E}/$\kappa$ of BaZrS$_{3}$ and its RP derivatives with a wide range of single crystals. The materials shown here range from soft, insulating crystals (e.g., Co$_6$S$_8$) to stiff, conductive crystals (e.g., diamond). The elastic moduli of BaZrS$_{3}$ and RP phases are significantly higher than those of other ultralow-thermal-conductivity single crystals (e.g., superatoms, metal halide perovskites, and layered perovskites) and are surpassed only by oxides and some semiconductors. Despite such strong bonding, the RP phases possess an ultralow thermal conductivity. As a result, the \textit{E}/$\kappa$ ratio of the Ba$_{3}$Zr$_{2}$S$_{7}$ single crystal is the highest reported to date. Though some polycrystalline materials have larger \textit{E}/$\kappa$ ratios,\cite{yang2019diffused,guo2022high} the record \textit{E}/$\kappa$ ratio shown in Figure 4 among single-crystalline solids is very significant in its own right as not only it eliminates any extrinsic spurious influences in \textit{E}/$\kappa$ values, but also allows us to study the fundamental mechanisms behind the “mechanically stiff phonon glass” paradigm. Additionally, the \textit{E}/$\kappa$ ratio of the RP phase is $\sim$3 times higher than that of BaZrS$_{3}$. Such \textit{E}/$\kappa$ difference between two phases of the same material is unprecedented in literature.

\begin{figure}[hbt!]
\begin{center}
\includegraphics[scale =0.35]{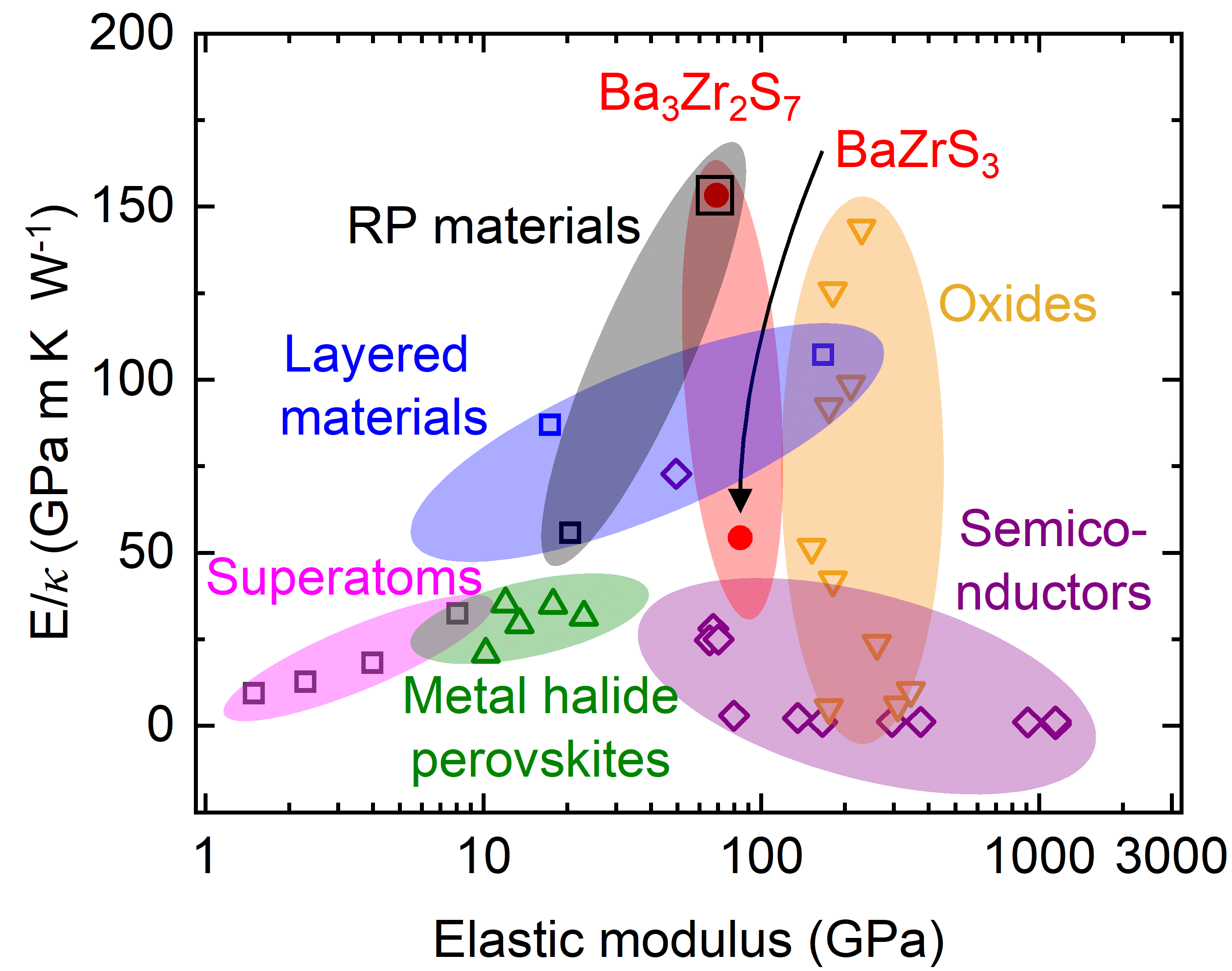}\\
\caption{Elastic modulus/thermal conductivity (\textit{E}/$\kappa$) ratio for a wide range of single-crystalline materials at room temperature. The crystals are grouped into superatoms, metal halide perovskites, semiconductors, oxides, layered materials, and BaZrS$_{3}$ and its RP derivatives. The thermal conductivity and elastic modulus data of the single crystals are provided in the Supporting Information.}   
\label{Figure 4}
\end{center}
\end{figure}

As shown in Figure 4, oxides generally possess a high elastic modulus. By replacing oxygen with sulfur, we reduced thermal conductivity significantly with moderate reduction in the elastic modulus. Introduction of strongly bonded periodic interfaces can further reduce the thermal conductivity without sacrificing stiffness proportionately. The sulfides studied here sit in an ideal regime of relatively high elastic modulus and low thermal conductivity thereby opening a paradigm for finding high \textit{E}/$\kappa$ ratio materials. For example, RP phases of BaHfS$_{3}$ are also likely to exhibit a high \textit{E}/$\kappa$ ratio. Moreover, the study of chalcogenide perovskites has gained momentum only recently. We anticipate the discovery and synthesis of a broad range of layered phases such as Aurivillius phases, Dion Jacobson phases in the near future.\cite{kendall1996recent,benedek2014origin} These layered materials will present additional opportunities to test and further the paradigm of “mechanically stiff phonon glass”.

\section{Conclusion}
The RP phases of BaZrS$_{3}$ is found to possess ultralow thermal conductivity and ultrahigh modulus-to-thermal conductivity ratio. We find that the rock-salt layers separating the perovskite sections of the RP structure lead to highly anisotropic thermal conductivity, with the cross-plane reaching values comparable to the amorphous solid despite similar and highly strong bonding across the full unit cell. Together with simulations, our results provide evidence that the rock-salt layers in the single crystal RP phases lead to ultralow phonon velocities, ultrashort phonon mean free paths, and strong localization within rock-salt layers, leading to ultralow, glass-like thermal conductivity. Our study provides a detailed overview of the mechanisms needed to achieve ultralow thermal conductivity in a non-vdW, strongly bonded, layered material.

\section{Acknowledgement}
We thank Kevin Ye and Rafael Jaramillo from Department of Materials Science and Engineering of Massachusetts Institute of Technology for their help with the sample preparation. Md.S.B.H., E.R.H., E.A.S., J.A.T., S.M., K.A. and P.E.H. appreciate support from the Office of Naval Research Grant No. N0014-23-1-2630. D.-L.B. and S.T.P. acknowledge support by the U.S. Department of Energy, Office of Science, Basic Energy Sciences, Materials Science and Technology Division Grant No. DE-FG02-09ER46554 and by the McMinn Endowment at Vanderbilt University. Computations were performed at the National Energy Research Scientific Computer Center (a U.S. Department of Energy Office of Science User Facility located at Lawrence Berkeley National Laboratory, operated under contract no. DE-AC02-05CH11231. B.Z., M.S., and J.R. acknowledge support from an ARO MURI program (W911NF-21-1-0327), an ARO grant (W911NF-19-1-0137), and National Science Foundation (DMR-2122071). H.Z. and T.F. acknowledge support from National Science Foundation (NSF) (award number: CBET 2212830). H.Z. and T.F. used the computational resource of Bridges-2 at Pittsburgh Supercomputing Center through allocation PHY220002 from the Advanced Cyber infrastructure Coordination Ecosystem: Services $\&$ Support (ACCESS) program, which is supported by National Science Foundation grants $\#$2138259, $\#$2138286, $\#$2138307, $\#$2137603, and $\#$2138296, National Energy Research Scientific Computing Center, a DOE Office of Science User Facility supported by the Office of Science of the U.S. Department of Energy under Contract No. DE-AC02-05CH11231 using NERSC award BES-ERCAP0022132, and Center for High Performance Computing (CHPC) at the University of Utah. A.G. and S.T. acknowledge support from the Office of Naval Research Grant No. N00014-21-1-2622 and National Science Foundation (NSF Award No. 2119365). This work was performed, in part, at the Center for Integrated Nanotechnologies, an Office of Science User Facility operated for the U.S. Department of Energy (DOE) Office of Science. Sandia National Laboratories is a multi-mission laboratory managed and operated by National Technology and Engineering Solutions of Sandia, LLC, a wholly owned subsidiary of Honeywell International, Inc., for the U.S. DOE’s National Nuclear Security Administration under contract DE-NA-0003525. The views expressed in the article do not necessarily represent the views of the U.S. DOE or the United States Government.

\newpage
\textbf{{{\Large Supporting Information}}}\\

\textbf{{{\Large S1. Growth details of BaZrS$_{3}$ and its RP derivatives}}}

The BaZrS$_{3}$ and Ba$_{3}$Zr$_{2}$S$_{7}$ single crystals are synthesized using the flux method, details of which can be found in previous publications.\cite{ye2022low} 1 g of BaCl$_2$ powder (Alfa Aesar, 99.998$\%$) is grounded and mixed with 0.5 g of stoichiometric mixtures of precursor powders (BaS, Zr, and S), and loaded into a quartz tube. For BaZrS$_{3}$, the tube is heated to 1050 $^\circ$C at a rate of 1.6 $^\circ$C/min, held at 1050 $^\circ$C for 100 h, cooled to 800 $^\circ$C at a rate of 0.1 $^\circ$C/min, and then to room temperature in an uncontrolled manner by shutting off the furnace. For Ba$_{4}$Zr$_{3}$S$_{10}$, the tube is heated to 1050 $^\circ$C at a rate of 0.3 $^\circ$C/min, held at 1050 $^\circ$C for 40 h, cooled to 400 $^\circ$C at a rate of 1 $^\circ$C/min, and then to room temperature in an uncontrolled manner. Ba$_{4}$Zr$_{3}$S$_{10}$ is formed as a secondary phase twined with some large Ba$_{3}$Zr$_{2}$S$_{7}$ crystals. X-ray diffraction along 001 orientation\cite{niu2019crystal} is used to characterize their existence, so Ba$_{4}$Zr$_{3}$S$_{10}$ layers are exfoliated to access them. The obtained samples are washed repeatedly with deionized water and isopropyl alcohol to remove excess flux before drying in airflow. The crystal dimensions are on the order of $\sim$100 $\mu$m. Amorphous BaZrS$_{3}$ thin films are grown from a phase pure BaZrS$_{3}$ target in a background gas mixture of hydrogen sulfide and argon (Ar-H$_2$S) on  LaAlO$_3$ substrate. The dense, polycrystalline target is prepared by sintering BaZrS$_{3}$ powders by high pressure torsion method.\cite{surendran2021epitaxial}\\

\textbf{{{\Large S2. Time-domain thermoreflectance (TDTR)}}}

We use a two-tint time-domain thermoreflectance (TDTR) setup to measure the thermal conductivity of the crystalline and amorphous BaZrS$_{3}$, Ba$_{3}$Zr$_{2}$S$_{7}$, and Ba$_{4}$Zr$_{3}$S$_{10}$ specimens.\cite{olson2020anisotropic,olson2020local} In our TDTR setup, a Ti:sapphire oscillator (80 MHz, $\sim$808 nm central wavelength, and $\sim$14 nm full width at half maximum) emanates subpicosecond laser pulses that are split into a high-power pump and a low-power probe beam. The pump beam is modulated at a frequency of 8.4 MHz by an electro-optic modulator (EOM) to create oscillatory heating events at the sample surface. The probe beam is then directed through a mechanical delay stage to detect the temporal change in thermoreflectivity which is related to the surface temperature change. Using a lock-in amplifier and a balanced photodetector, the probe beam measures the temperature decay up to 5.5 ns. The TDTR data are analyzed by fitting a cylindrically symmetric, multilayer thermal model to the ratio of in-phase to out-of-phase signal (--\textit{V}$\sous{\footnotesize in}$/\textit{V}$\sous{\footnotesize out}$) from the RF lock-in amplifier.\cite{cahill2002thermometry,cahill2004analysis,schmidt2008pulse,hopkins2010criteria} Figure S1(a) shows the best-fit thermal model to
the TDTR data for the BaZrS$_{3}$, Ba$_{3}$Zr$_{2}$S$_{7}$, and Ba$_{4}$Zr$_{3}$S$_{10}$ single crystals. For all the measurements, the pump and probe beams are coaxially focused to $\sim$20 and 11 $\mu$m 1/e$^2$ diameters, respectively on the sample surface. Prior to the measurements, all the samples are coated with an $\sim$80 nm aluminum film via electron beam evaporation for optothermal transduction.\cite{decoster2019thermal} Our TDTR setup is calibrated with a sapphire and a high purity fused silica wafer.    

In TDTR, we fit for the thermal conductivity of the specimen and thermal boundary conductance between the aluminum transducer and specimen. Figure S1(b) shows the TDTR sensitivity calculations for the BaZrS$_{3}$ crystal. The sensitivity analysis is performed following the methodology published in prior literature.\cite{koh2009comparison,jiang2016accurate,giri2016heat} As exhibited here, TDTR measurements have low sensitivity to the thermal boundary conductance and high sensitivity to the BaZrS$_{3}$ thermal conductivity. TDTR measurements also have high sensitivity to the volumetric heat capacity of the aluminum transducer and BaZrS$_{3}$ crystals. \\

\begin{figure}[hbt!]
\begin{center}
\renewcommand{\thefigure}{S1}
\includegraphics[scale = 0.36]{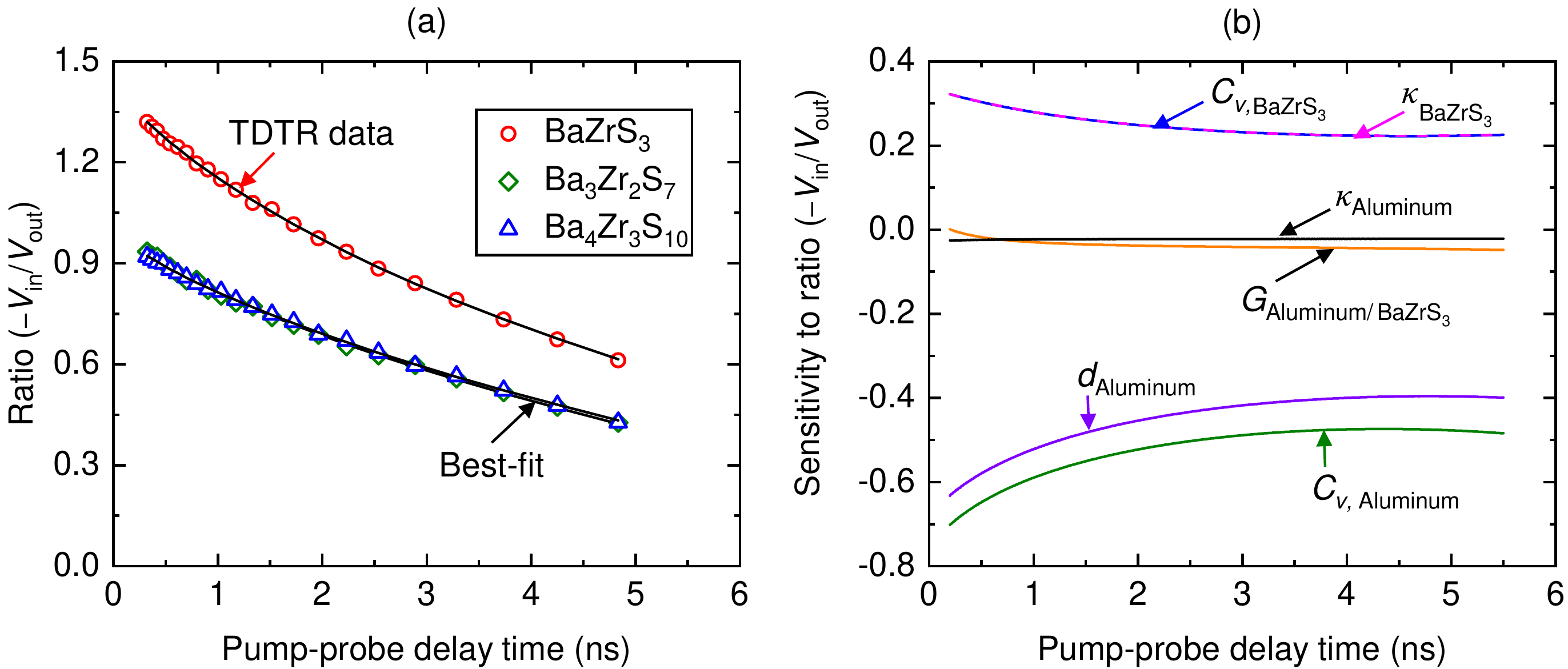}\\
\caption{(a) Best-fit thermal model to the TDTR data for BaZrS$_{3}$, Ba$_{3}$Zr$_{2}$S$_{7}$, and Ba$_{4}$Zr$_{3}$S$_{10}$ crystals at room temperature. (b) Sensitivity of TDTR measurements to the ratio of in-phase to out-of-phase signal (--\textit{V}$\sous{\footnotesize in}$/\textit{V}$\sous{\footnotesize out}$) for BaZrS$_{3}$ at room temperature. Here, \textit{C}$\sous{\footnotesize v}$, $\kappa$, \textit{G}, and \textit{d} represent volumetric heat capacity, thermal conductivity, thermal boundary conductance, and thickness, respectively.}
\label{Figure 2}
\end{center}
\end{figure}

\newpage
\textbf{{{\Large S3. Steady-state temperature rise during TDTR measurements}}}

Since the RP phases possess ultralow thermal conductivity, steady-state temperature rise from pulsed laser heating can be a major concern, particularly at low temperatures. Therefore, at each temperature, we calculate the steady-state temperature rise for the precise aluminum/sample geometry from the numerical solution to the cylindrical heat equation.\cite{braun2018steady} For each sample, we optimize the laser power to obtain good signal to noise ratio and minimal steady-state temperature rise. The maximum steady-state temperature rise observed in our study is 11 K. This degree of steady-state temperature rise is not expected to change the properties of aluminum and samples significantly to cause any error.\cite{sun2020high}\\

\textbf{{{\Large S4. Uncertainty analysis}}}

We calculate the uncertainty ($\triangle$) of TDTR measurements using the following equation\cite{braun2019steady,wei2013invited,yang2016uncertainty}

\begin{equation} 
\renewcommand{\theequation}{S1}
\triangle = \sqrt{(\sigma)^{2} + \bigg(R \cdot \frac{{\delta}_{\phi}}{{S}_{\kappa}}\bigg)^{2} + \sum \bigg(\frac{{S}_{\alpha}}{{S}_{\kappa}} \cdot \frac{{\delta}_{\alpha}}{{\alpha}}\bigg)^{2}}
\label{eq4}
\end{equation}

\noindent here the first term $\sigma$ represents the standard deviation of multiple measurements across different spots. The second term \textit{R}$\frac{{\delta}_{\phi}}{{S}_{\kappa}}$ represents the uncertainty in determining the absolute value of the phase from the lock-in amplifier. \textit{R}, ${{\delta}_{\phi}}$, and ${{S}_{\kappa}}$  stand for the ratio of in-phase to out-of-phase signal, uncertainty of the phase, and sensitivity of \textit{R} to $\kappa$ in the thermal model, respectively. The third term $\frac{{S}_{\alpha}}{{S}_{\kappa}}$$\frac{{\delta}_{\alpha}}{{\alpha}}$ represents the propagation of uncertainty from individual input parameters to the thermal model. ${{\delta}_{\alpha}}$ and ${{S}_{\alpha}}$ are uncertainty of parameter $\alpha$ and sensitivity of \textit{R} to $\alpha$ in the thermal model, respectively. 

We conduct the TDTR measurements at 8.4 MHz modulation frequency. At such high modulation frequency, the signal noise and ${\delta}_{\phi}$ is quite small.\cite{koh2009comparison} Therefore, the main sources of uncertainty in the TDTR measurements are the standard deviation and input parameters. 

The four input parameters to the thermal model are: volumetric heat capacity of aluminum and samples, aluminum transducer thickness, and aluminum thermal conductivity. The volumetric heat capacities of aluminum and samples are adopted from literature\cite{touloukian1970specific} and density functional theory (DFT) calculations, respectively. The thickness and thermal conductivity of the aluminum transducer are measured by picosecond acoustics\cite{o2001characterization} and four-point probe using the Wiedemann–Franz law\cite{park2018direct}, respectively.

Figure S2(a)-(d) shows the contour uncertainty analysis for various input parameters. The dashed lines represent the 95$\%$ confidence interval for the thermal conductivity corresponding to a specific value of the input parameter. This confidence interval confirms the uncertainty bounds of thermal conductivity for the parameters. Details of the contour uncertainty analysis can be found in previous publications.\cite{koh2020thermal}

\begin{figure}[hbt!]
\begin{center}
\renewcommand{\thefigure}{S2}
\includegraphics[scale = 0.36]{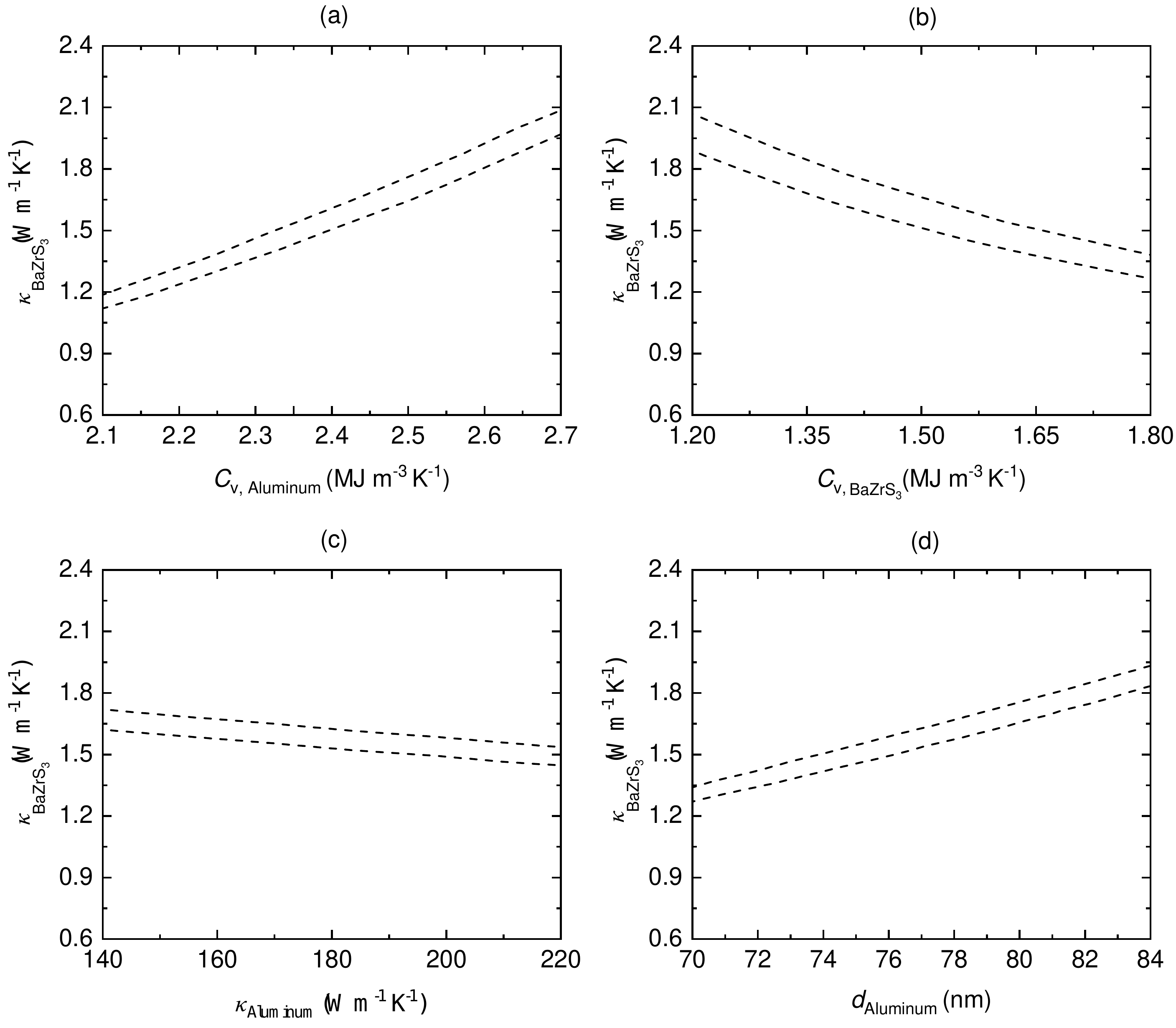}\\
\caption{Uncertainty contour analysis of TDTR measurements showing a combination of BaZrS$_{3}$ thermal conductivity and (a) volumetric heat capacity of aluminum (2.42 MJ m$^{-3}$ K$^{-1}$), (b) volumetric heat capacity of BaZrS$_{3}$ (1.505 MJ m$^{-3}$ K$^{-1}$), (c) aluminum thermal conductivity (180 W m$^{-1}$ K$^{-1}$), and (d) aluminum thickness (77 nm).}
\label{Figure 2}
\end{center}
\end{figure}

\newpage
\textbf{{{\Large S5. Steady-state thermoreflectance (SSTR)}}}

Along with TDTR, we employ another optical pump-probe technique named steady-state thermoreflectance (SSTR) to measure the thermal conductivity. Unlike TDTR, SSTR technique is insensitive to the volumetric heat capacity of aluminum and samples. We employ this technique to validate the volumetric heat capacity assumed for amorphous BaZrS$_{3}$ film. In SSTR setup, a continuous wave pump laser (532 nm wavelength) is modulated at 100 Hz frequency to create steady-state temperature rise at the sample surface. The reflectivity change and temperature rise are detected by a continuous wave probe laser (786 nm wavelength). By changing the pump laser power, the heat flux deposited on the sample surface and corresponding steady-state temperature rise can be changed. By correlating the pump laser power and temperature rise via Fourier's law, the thermal conductivity of any material can be derived. More details regarding the SSTR setup can be found in previous publications.\cite{braun2019steady,hoque2021high,hoque2021thermal} We use 1/e$^2$ pump and probe diameters of $\sim$20 $\mu$m for the measurements.\\

\textbf{{{\Large S6. Volumetric heat capacity of BaZrS$_{3}$, Ba$_{3}$Zr$_{2}$S$_{7}$, and Ba$_{4}$Zr$_{3}$S$_{10}$}}}

The volumetric heat capacity of crystalline BaZrS$_{3}$ and Ba$_{3}$Zr$_{2}$S$_{7}$ are calculated using DFT. Within the framework of DFT, the projector augmented wave method (PAW)\cite{kresse1999ultrasoft} as implemented in the Vienna ab initio simulation package (VASP)\cite{hafner2008ab} code is employed for relaxing the cell parameter and atomic positions of all perovskite unit cells. The PBEsol\cite{perdew2008restoring} functional for solids is utilized for relaxing the unit cell and atomic positions. The Methfessel-Paxton\cite{methfessel1989high} smearing scheme with the gamma parameter is set to 0.1 eV and an energy cut-off of 500 eV is used for the planewaves expansion. A 5 $\times$ 5 $\times$ 1 Monkhorst-Pack\cite{monkhorst1976special} special grid sampling of the k-points is used for integration in the Brillouin zone for structural optimization. For resolution of the Kohn-Sham equations, the self-consistent field procedure is considered by setting energy changes for each cycle at 10$^{-5}$ eV as the convergence criterion between two successive iterations. The atomic forces are calculated by minimizing the total forces until the energy convergence is less than 10$^{-3}$ eV. The interatomic force constants (IFCs) are computed by utilizing density functional perturbation theory with a 2 $\times$ 2 $\times$ 1 supercell including up to the seventh nearest neighbor interactions. The PHONOPY\cite{togo2015first} code is used to obtain the harmonic displacements while considering all neighboring interactions. For the anharmonic interactions, a 2 $\times$ 2 $\times$ 1 supercell with up to the seventh neighbor interactions is employed with the THIRDORDER.PY code.\cite{carrete2017almabte,li2012thermal} 

Figure S3 shows the volumetric heat capacity of crystalline BaZrS$_{3}$ and Ba$_{3}$Zr$_{2}$S$_{7}$. We assume that the crystalline and amorphous BaZrS$_{3}$ possess the same volumetric heat capacity. To ensure the validity of this assumption, we measure the room-temperature thermal conductivity of a $\sim$107 nm amorphous BaZrS$_{3}$ film grown on silicon substrate via both TDTR and SSTR techniques. The TDTR technique is sensitive to heat capacity whereas the SSTR is insensitive to it. The measured thermal conductivity is in excellent agreement between the two techniques, thus proving the accuracy of the volumetric heat capacity assumed for amorphous BaZrS$_{3}$ in TDTR. We further assume that the Ba$_{3}$Zr$_{2}$S$_{7}$ and Ba$_{4}$Zr$_{3}$S$_{10}$ possess the same volumetric heat capacity. A 10$\%$ uncertainty is used with all the heat capacity values in TDTR measurements.\\

\begin{figure}[hbt!]
\begin{center}
\renewcommand{\thefigure}{S3}
\includegraphics[scale = 0.51]{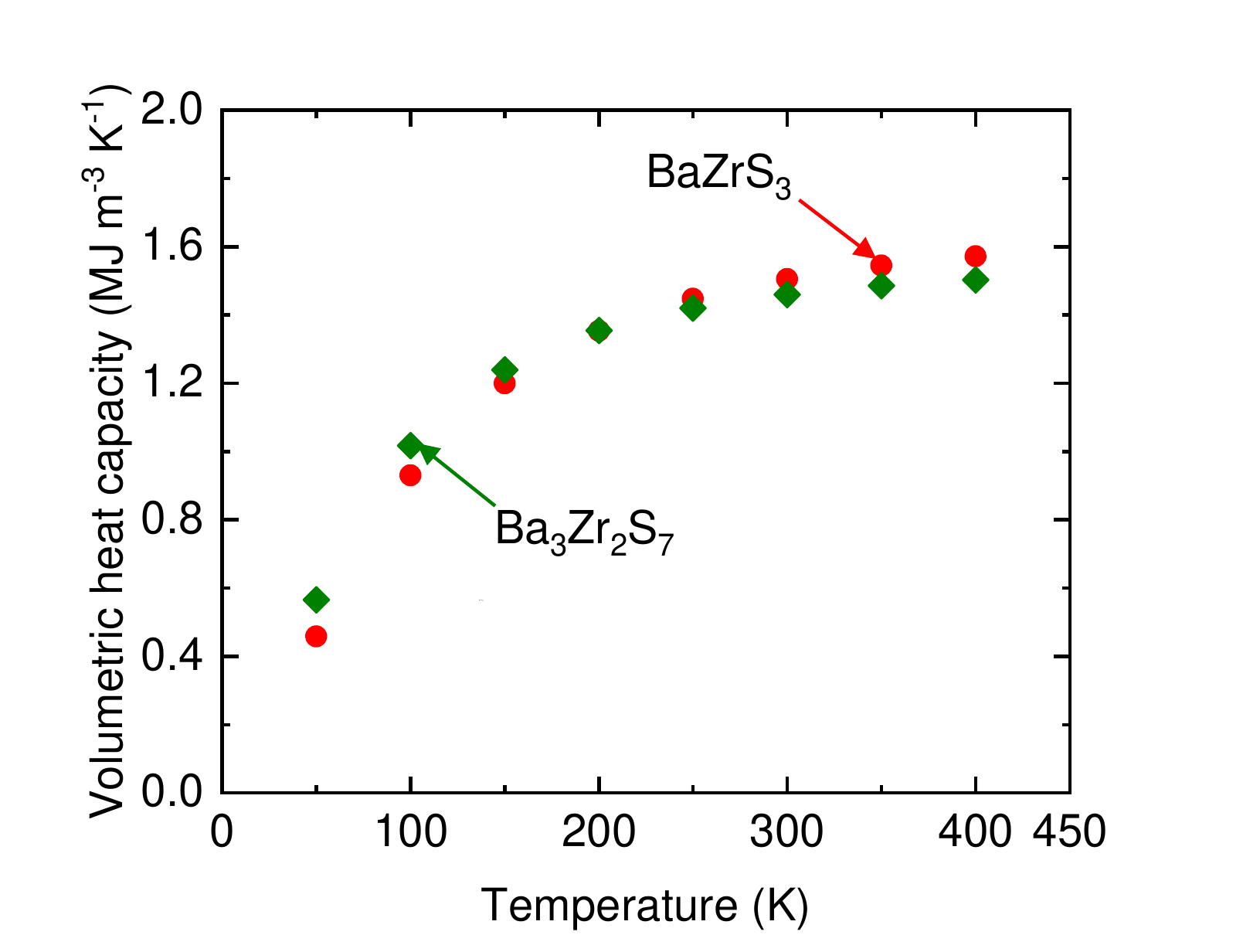}\\
\caption{Volumetric heat capacity of crystalline BaZrS$_{3}$ and Ba$_{3}$Zr$_{2}$S$_{7}$ calculated via DFT.}
\label{Figure 2}
\end{center}
\end{figure}

\newpage
\textbf{{{\Large S7. Ion irradiation of the BaZrS$_{3}$ and Ba$_{4}$Zr$_{3}$S$_{10}$ crystals}}}

The BaZrS$_{3}$ and Ba$_{4}$Zr$_{3}$S$_{10}$ single crystals are irradiated with gold (Au) ions at an energy of 2.8 MeV using a 6 MV tandem Van de Graaff accelerator. The ion implantation depths are calculated via SRIM simulations for an ion energy of 2.8 MeV. Details of the SRIM simulations for determining the stopping range of ions can be found in previous publications.\cite{scott2021reductions,scott2021thermal} The implantation depths of the Au ions are greater than 450 nm. This length scale is much larger than the thermal penetration depth of TDTR measurements.\cite{koh2007frequency} Therefore, the thermally probed region and the measured thermal conductivity are of the defected region pre-end-of-range.\cite{scott2020orders,scott2021probing}

For the irradiation, six crystals of each material are carbon-taped onto a silicon substrate and loaded into the implant chamber. Afterwards, the chamber is pumped down to a pressure on the order of 10$^{-7}$ torr. Due to the small size of the crystals  ($\sim$100 $\times$ 100 $\times$ 100 $\mu$m$^3$) relative to the incident ion beam, spatial uniformity is achieved during the implantation. Nominal fluences ranging from 1.6 $\times$ 10$^{11}$ to 8 $\times$ 10$^{13}$ cm$^{-2}$ are applied to each material. 

\begin{figure}[hbt!]
\begin{center}
\renewcommand{\thefigure}{S4}
\includegraphics[scale = 0.51]{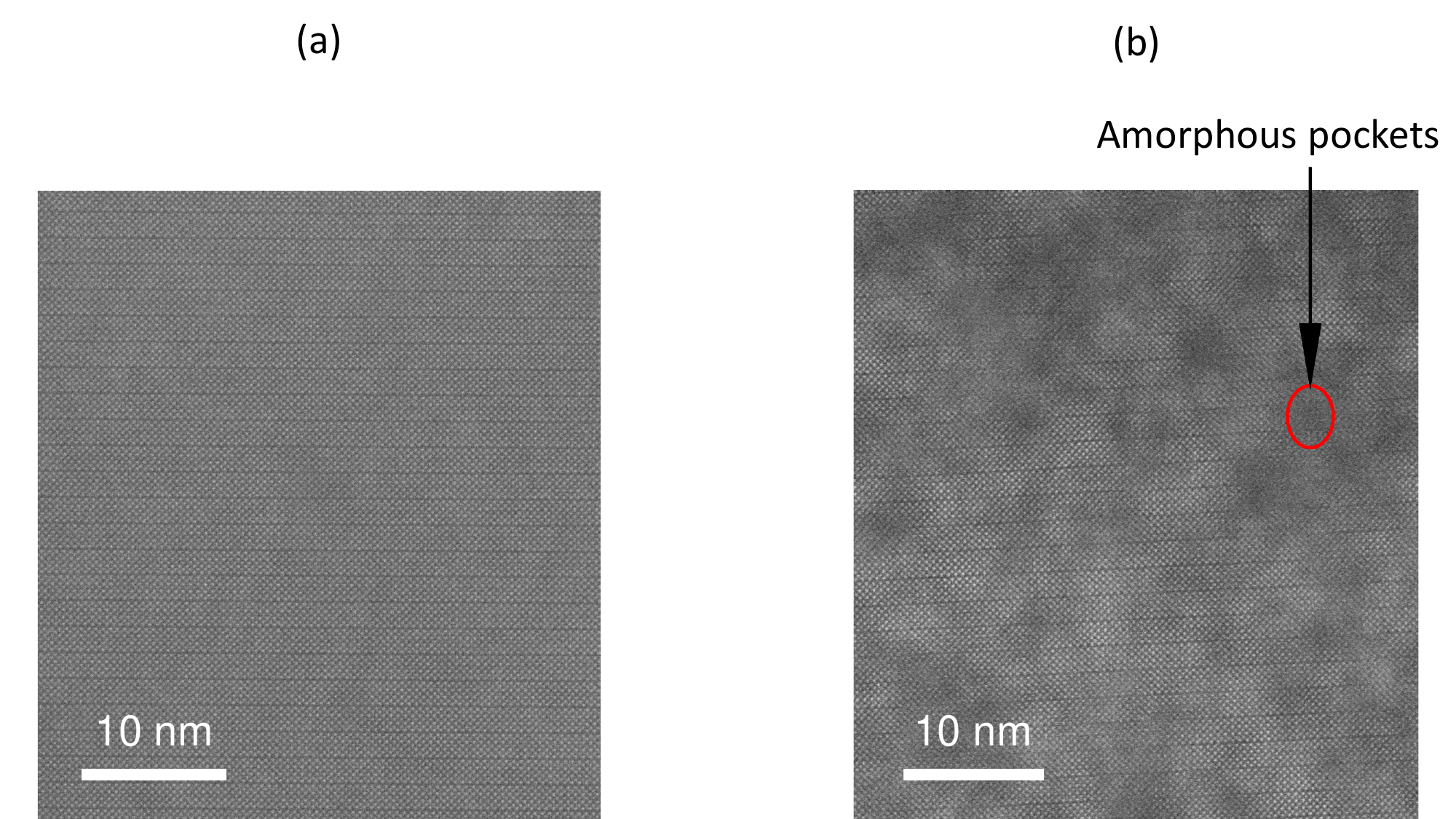}\\
\caption{TEM images of Ba$_{4}$Zr$_{3}$S$_{10}$ crystals corresponding to (a) 1.6 $\times$ 10$^{11}$ and (b) 4.8 $\times$ 10$^{13}$ cm$^{-2}$ ion doses.}
\label{Figure 2}
\end{center}
\end{figure}

\newpage
TEM images of irradiated Ba$_{4}$Zr$_{3}$S$_{10}$ crystals are shown in Figure S4. As exhibited here, when the ion dose is 1.6 $\times$ 10$^{11}$ cm$^{-2}$, no visible damage is observed in the crystal. However, as the ion dose is increased to 4.8 $\times$ 10$^{13}$ cm$^{-2}$, amorphous pockets are introduced in the system. For both ion doses, the layerings of Ba$_{4}$Zr$_{3}$S$_{10}$ remain uninterrupted. This verifies the strong bonding across the rock-salt layers in the RP phases and makes them suitable for deep space applications in radiation environments.\\

\textbf{{{\Large S8. Presence of nano-domains in BaZrS$_{3}$ crystals}}}

One of the biggest requirements for any thermoreflectance measurement is optically smooth sample surface. To obtain such surface quality, we mechanically polish the samples. Polishing deforms the crystals and creates nano-scale domains in BaZrS$_{3}$ as exhibited in Figure S5(a). The presence of such nano-domains can obfuscate the origin of thermal behaviors in BaZrS$_{3}$. Therefore, as an alternate to polishing, we use cleaving on several BaZrS$_{3}$ crystals. Although cleaving does not deform the material and create nano-domains, obtaining large quantities of optically smooth crystals can be highly challenging via cleaving. As a result, we use polished BaZrS$_{3}$ crystals for many of our measurements. The BaZrS$_{3}$ thermal conductivity shown in Figure 2(a) corresponds to a cleaved crystal, whereas the ones shown in Figure 2(c) belong to polished crystals. The cleaved and polished BaZrS$_{3}$ crystals have thermal conductivities of 1.55 $\pm$ 0.2 and 1.21 $\pm$ 0.18 W m$^{-1}$ K$^{-1}$, respectively, at room temperature. 

Polishing also deforms the RP crystals and creates domains. However, due to the high stiffness of the RP phases, the domain sizes are greater than 500 nm as shown in Figure S5(b). At such length scales, the domains are not expected to impact the thermal conductivity of the RP phases.\cite{osei2019understanding} As a result, we use mechanically polished RP crystals for all of our measurements.\\

The thermal conductivity of the cleaved and polished BaZrS$_{3}$ crystals are shown in Figure S5(c) as a function of temperature. Due to phonon scattering at the domain boundary, the thermal conductivity of the polished crystal is lower than the cleaved one throughout the temperature range.

\begin{figure}[hbt!]
\begin{center}
\renewcommand{\thefigure}{S5}
\includegraphics[scale = 0.51]{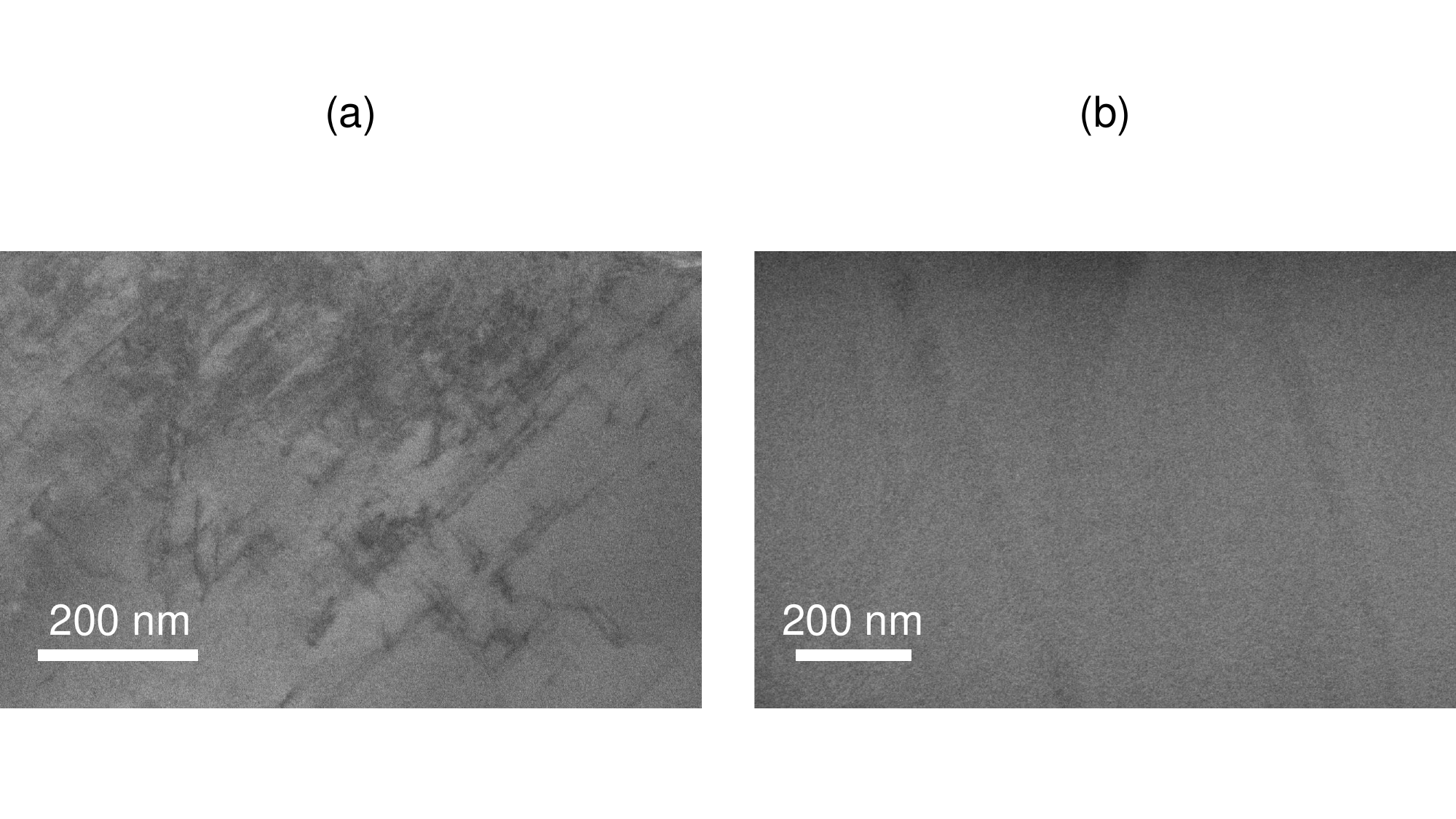}\hfill
\includegraphics[scale = 0.51]{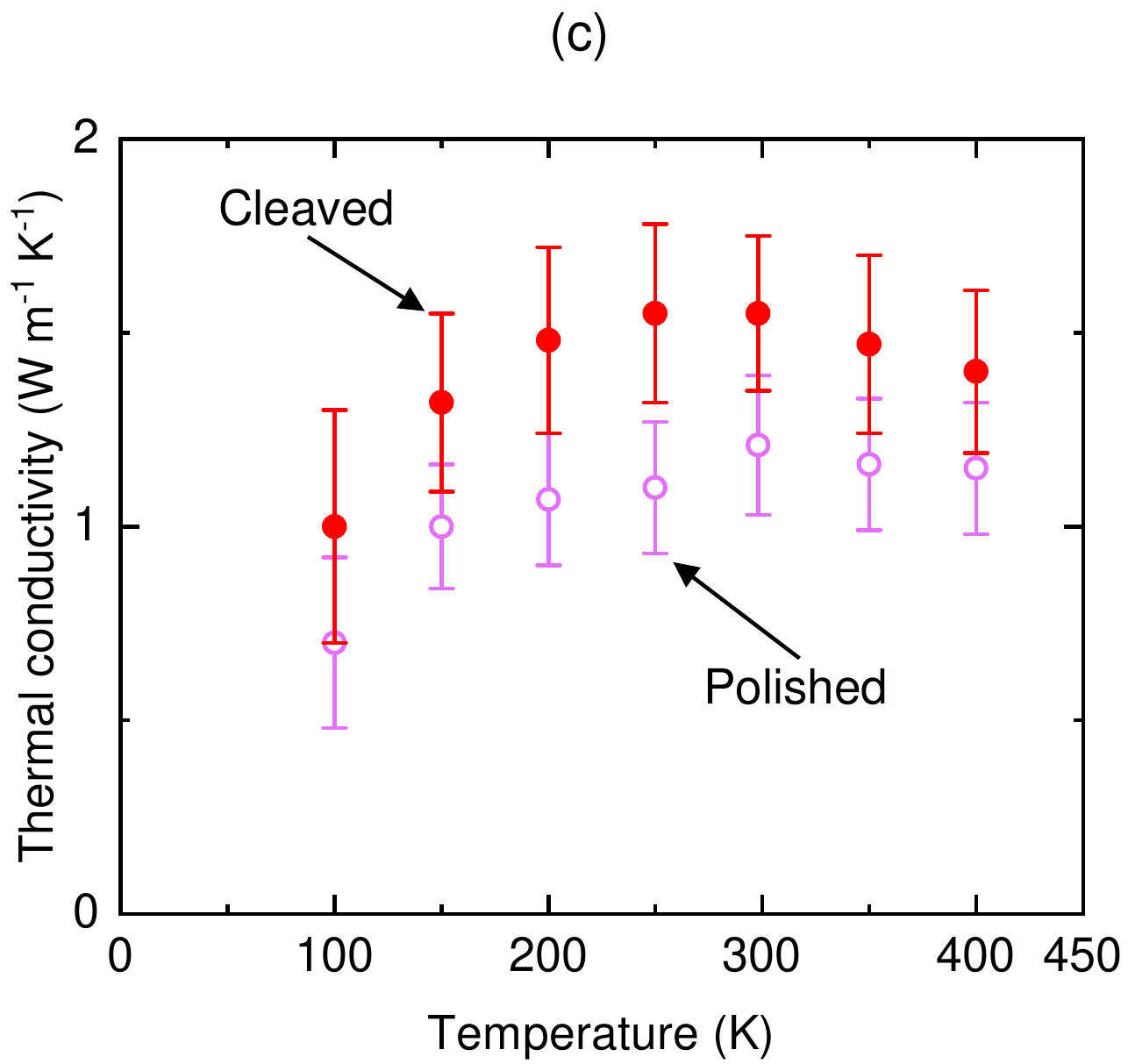}\hfill
\caption{TEM images of domains in (a) BaZrS$_{3}$ and (b) Ba$_{3}$Zr$_{2}$S$_{7}$ crystals. (c) Thermal conductivity of cleaved and polished BaZrS$_{3}$ crystals as a function of temperature.}
\label{Figure 2}
\end{center}
\end{figure}

\newpage
\textbf{{{\Large S9. Machine learning molecular dynamics (MLMD) simulation details}}}

To construct the training database of machine learning potential, first principles calculations of both BaZrS$_{3}$ and Ba$_{3}$Zr$_{2}$S$_{7}$ are performed by using VASP\cite{kresse1996efficient} with the PAW\cite{blochl1994projector} method based on the DFT. Local density approximation (LDA)\cite{perdew1981self} is chosen as the exchange-correlation functional. The plane-wave energy cutoff is selected as 500 eV. The primitive cell is first relaxed with the energy convergence threshold of 10$^{-8}$ eV and the force convergence threshold of 10$^{-4}$ eV/Å between atoms. For BaZrS$_{3}$, the calculational primitive cell contains 20 atoms following the \textit{Pnmb} symmetry, where the lattice constants are 7.03 Å, 9.86 Å, and 6.89 Å along \textit{a}, \textit{b}, and \textit{c} directions, respectively. The \textbf{k}-mesh is set to 7 × 5 × 7. For Ba$_{3}$Zr$_{2}$S$_{7}$, the calculational primitive cell contains 24 atoms following the \textit{I4mmm} symmetry, where the lattice constants are 4.92 Å, 4.92 Å, and 25.24 Å along \textit{a}, \textit{b}, and \textit{c} directions, respectively. The \textbf{k}-mesh is set to 10 × 10 × 2. After the relaxation, the VASP + PHONOPY combination is utilized to calculate the phonon properties of BaZrS$_{3}$ and Ba$_{3}$Zr$_{2}$S$_{7}$. \textit{Ab initio} molecular dynamics (AIMD) simulations are performed to sample the potential energy surface. For BaZrS$_{3}$ and Ba$_{3}$Zr$_{2}$S$_{7}$, 2 × 2 × 2 supercells are adopted containing 160 and 192 atoms, respectively, in the simulation domains. The time step of AIMD is set to 5 fs and a total step of 8000 is run for each individual AIMD. Independent AIMDs at 100 K, 200 K, 300 K, 400 K, and 500 K with Canonical ensemble (NVT) are performed. During the AIMD process, energies, forces, and stresses are recorded together with corresponding atomic configurations to form the training database for machine learning potential.

The potential energy surface is approximated by moment tensor potential (MTP) trained by Machine-Learning Interatomic Potentials (MLIP) package.\cite{novikov2020mlip} MTP represents the energy of an atomic configuration cfg (\textit{E}$^{\text{mtp}}$(cfg)) as a sum of contributions of local atomic environments of each atom, denoted by \textit{n$_i$}, shown as

\begin{equation} 
\renewcommand{\theequation}{S2}
E^{\text{mtp}}(\text{cfg})  = \sum_{i=1}^{n}V(n_i) 
\label{eq1}
\end{equation}

MTP with different levels are provided and a higher level means more fitting parameters, potential higher accuracy and possible issues of overfitting. Given a training database containing K configurations cfg$_k$ (k=1, 2, …, K) with their corresponding DFT calculated energies, forces, and stresses, the training process of MTP is minimizing the following expression:

\begin{equation} 
\renewcommand{\theequation}{}
\sum_{k=1}^{K}\bigg[w_{e}\bigg({E}^{\text{mtp}}(\text{cfg}_k) - {E}^{\text{DFT}}(\text{cfg}_k)\bigg)^2 +  w_{f}\sum_{k=1}^{N_k}\bigg(\text{f}_{i}^{\text{mtp}}(\text{cfg}_k) - \text{f}_{i}^{\text{DFT}}(\text{cfg}_k)\bigg)^2 
+ w_{s}\bigg({\sigma}^{\text{mtp}}(\text{cfg}_k) - {\sigma}^{\text{DFT}}(\text{cfg}_k)\bigg)^2\bigg]
\nonumber
\end{equation}

\noindent where \textit{N$_k$} is the number of atoms in the \textit{k}th configuration, \textit{w$_e$}, \textit{w$_f$}, and \textit{w$_s$} are non-negative weights. More details about MTP and the training process can be found in Novikov \textit{et al.}\cite{novikov2020mlip} In this work, MTP with the level of 22 is adopted to fit the potential energy surface and the training step is set to be 20,000.

Once the MTP is developed, equilibrium molecular dynamics (EMD) is performed via LAMMPS\cite{kryuchkov2018complex} package to get the lattice thermal conductivity, $\kappa$. A supercell of 10 × 10 × 10 conventional cell is adopted for BaZrS$_{3}$ and 10 × 10 × 10 for Ba$_{3}$Zr$_{2}$S$_{7}$, respectively, which should be large enough to avoid size effect.\cite{wang2017uncertainty} The time step of EMD is 5 fs, and periodic boundary conditions are implemented in all three directions. The simulation is carried out first by 2 ns NVT, followed by a 2 ns NVE to fully relax the lattice, and then another NVE of 4 ns, during which the heat current \textbf{\textit{J}} is recorded. Based on the Kubo-formula,\cite{green1954markoff,kubo1957statistical} $\kappa$ along a certain direction $\alpha$ is proportional to the integral of autocorrelation of heat current:

\begin{equation} 
\renewcommand{\theequation}{S3}
\kappa{_\alpha}  = \frac{1}{k{_B}VT^2}\int_{0}^{\infty}(J_\alpha(0) \cdot J_\alpha(t))dt 
\label{eq1}
\end{equation}

\noindent where \textit{k$_{B}$} is Boltzmann constant, \textit{V} is the volume of the simulation domain, \textit{T} is temperature, and \textit{t} is time. In LAMMPS, the heat current vector \textbf{\textit{J}} is defined as

\begin{equation} 
\renewcommand{\theequation}{S4}
\textbf{\textit{J}}  = \sum_{i}^{}v_iE_i -\sum_{i}^{}\textbf{\textit{S}}_iv_i = \sum_{i}^{}v_iE_i + \sum_{i<j}^{}(\textbf{\textit{F}}_{ij} \cdot v_i)\textbf{\textit{r}}_{ij}
\label{eq1}
\end{equation}

\noindent where \textit{E$_i$} is the total energy of the \textit{i}th atom, \textit{v$_i$} is the velocity vector of the \textit{i}th atom, \textbf{\textit{F}}$_{ij}$ is the force interaction between the \textit{i}th and \textit{j}th atom, and \textbf{\textit{r}}$_{ij}$ represents the position vector between the \textit{i}th and \textit{j}th atom. The final $\kappa$ is averaged over 8 independent EMD simulations at each temperature with different initial velocities. Quantum correction is taken into account due to the classical specific heat used in LAMMPS is much higher than the real specific heat at low temperature.\cite{turney2009assessing}\\

\textbf{{{\Large S10. Spectral energy density (SED) calculations}}}

To gain insights into the intrinsic mechanisms that dictate the thermal transport in BaZrS$_{3}$ and its RP phase Ba$_{3}$Zr$_{2}$S$_{7}$ structures, we calculate the phonon mode specific properties by performing spectral energy density (SED) calculations in MD simulations, which is given as\cite{thomas2010predicting,thomas2015erratum}

\begin{equation} 
\renewcommand{\theequation}{S5}
\Phi(\mathbf{q},\omega)  = \frac{1}{4\pi\tau N{_T}} {\sum_{\alpha}^{3}}{\sum_{b}^{B}}m_{b}\bigg|\int_{0}^{\tau} \sum_{n_{x,y,z}}^{N_T}\dot{u}_{\alpha}\bigg(\begin{matrix}n_{x,y,z}\\b\end{matrix};t\bigg)\times exp\bigg[i\mathbf{q\cdot r}\bigg(\begin{matrix}n_{x,y,z}\\0\end{matrix}\bigg)-i\omega t\bigg]dt\bigg|^2 
\label{eq1}
\end{equation}

\noindent where $\tau$ is the total simulation time, $\alpha$ is the cartesian direction, \textit{n$_{x,y,z}$} is a unit cell, \textit{N$_{T}$} is the number of unit cells in the crystal, \textit{b} is the atom label in a given unit cell, \textit{B} is the atomic number in the unit cell, \textit{m$_{b}$} is the mass of atom \textit{b} in the unit cell, $\dot{u}_{\alpha}$ denotes the velocity along the $\alpha$ direction at time \textit{t}, and $\mathbf{r}$ is the equilibrium position of each unit cell.

To ensure a high resolution in our SED calculations, we create the computational domain for both BaZrS$_{3}$ and its RP phase Ba$_{3}$Zr$_{2}$S$_{7}$ structures with \textit{N$_{T}$} = 400 (representing the total number
of unit cells) and perform SED calculations for 100 $\mathbf{q}$-points. Initially, we equilibriate our supercell structure using the Nosé-Hoover thermostat and barostat\cite{hoover1985canonical} for 1 ns with a time step of 0.5 fs where the number of particles, pressure, and temperature of the system are held constant at ambient pressure. After the NPT integration, we further equilibriate our structures in the NVT ensemble, keeping the volume and temperature constant for an additional 1 ns. Finally, for the data collection for our SED calculations, we record the velocities and positions of each atom in the microcanonical ensemble (or NVE ensemble) for a total of 1.5 ns.

\clearpage
\textbf{{{\Large S11. Elastic modulus measurements and derivation of sound speed}}}

We measure the elastic modulus of the BaZrS$_{3}$ crystals using a nanoindeter (MTS XP) based on the standard continuous stiffness measurement (CSM) method.\cite{oliver1992improved} Before the measurements, the nanoindenter is calibrated with a silica standard. Details of the calibration procedure can be found in previous publications.\cite{ding2020thermal,braun2018charge} The elastic modulus of BaZrS$_{3}$ is determined to be 84.2 $\pm$ 5.4 GPa. This value is in agreement with literature.\cite{zhang2020first} 

The elastic modulus of Ba$_{3}$Zr$_{2}$S$_{7}$ is measured by a Bruker Hysitron TI 950 Triboindenter. The indenter is calibrated with a single crystal quartz sample with known modulus and hardness (69.6 GPa and 9.25 GPa, respectively). The measured elastic modulus of Ba$_{3}$Zr$_{2}$S$_{7}$ is 69 $\pm$ 4.5 GPa. The indentation curves exhibit pop-in features, which are sudden displacement excursions; a common feature observed in single crystals.\cite{xia2016single}  

Assuming isotropic elastic properties, the sound speed components of BaZrS$_{3}$ can be derived using the following equations\cite{braun2016breaking}

\begin{equation} 
\renewcommand{\theequation}{S6}
v_{L} = \sqrt{\frac{E(1 - \nu)}{\rho (1 + \nu)(1 - 2\nu)}} 
\label{eq4}
\end{equation}

\begin{equation} 
\renewcommand{\theequation}{S7}
v_{T} = \sqrt{\frac{E}{2\rho (1 + \nu)}} 
\label{eq4}
\end{equation}

\begin{equation} 
\renewcommand{\theequation}{S8}
v_{s} = \frac{1}{3}(2v_{T} + v_{L}) 
\label{eq4}
\end{equation}

\noindent where \textit{v}$_{L}$, \textit{v}$_{T}$, \textit{v}$_{s}$, \textit{E}, $\rho$, and $\nu$ represent longitudinal sound speed, transverse sound speed, sound velocity, elastic modulus, density, and Poisson’s ratio, respectively. We take the density and Poisson's ratio to be 4.39 g cm$^{-3}$ and 0.28 $\pm$ 0.05, respectively.\cite{niu2019crystal} The derived sound speed components of BaZrS$_{3}$ are tabulated Table S1. The error bars incorporate the uncertainty of the elastic modulus and Poisson's ratio.

To further verify the longitudinal sound speed, we use picosecond acoustics on a 416 nm crystalline BaZrS$_{3}$ film. Details of our picosecond acoustics metrology and data reduction can be found in previous publications.\cite{gorham2014density,braun2016breaking,hoque2023interface} Picosecond acoustics reveal the longitudinal sound speed of BaZrS$_{3}$ to be 4818 $\pm$ 512 m s$^{-1}$, in excellent agreement with the elastic modulus derived value. The error bar of the picosecond acoustic measurement incorporates the standard deviation and uncertainty of the film thickness ($\sim$44 nm).\\  

\begin{table*}[hbt!]
\renewcommand{\thetable}{S1}
\centering
\caption{Longitudinal sound speed, transverse sound speed, and sound velocity of the BaZrS$_{3}$ crsytals.}
\begin{tabular}{cc}
\hline\\
Sound speed components & Velocity (m s$^{-1}$)
\vspace{3mm}\\
\hline\\
\textit{v}$_{L}$ & 4952 $\pm$ 410
\vspace{1mm}\\
\textit{v}$_{T}$ & 2737 $\pm$ 112
\vspace{1mm}\\
\textit{v}$_{s}$ & 3475 $\pm$ 143
\vspace{1mm}\\
\hline
\end{tabular}
\end{table*}

\clearpage
\textbf{{{\Large S12. Elastic modulus calculations}}}

To determine the stress-strain relationship for our BaZrS$_3$ and its RP phase Ba$_3$Zr$_2$S$_7$ structures, we use MLMD simulations to calculate the elastic moduli. We deform the simulation cell in the uniaxial direction using a strain rate of 10$^{8}$ s$^{-1}$. Concurrently, we maintain zero pressure at the boundaries in the remaining two lateral directions. This process is performed under the NPT ensemble, where the system's number of particles, pressure, and temperature are kept constant. The strain is computed at intervals of 0.5 fs using the formula $(L_x-L)/L$, where $L_x$ represents the current length and $L$ denotes the initial length of the domain in the direction of the applied strain. Simultaneously, the stress of the entire structure is recorded every 50 fs to enable the generation of a stress-strain relationship for the structures.

For BaZrS$_3$, we calculate a Young's modulus value of 85 GPa in the cross-plane direction. For Ba$_3$Zr$_2$S$_7$, we calculate Young's modulus values of 64.5 GPa and 64.3 GPa in the in-plane directions, and 81.3 GPa in the cross-plane direction.

\newpage
\textbf{{{\Large S13. Minimum limit and diffuson limit}}}

Cahill \textit{et al.}'s minimum limit model states that the lower bound of thermal conductivity can be achieved when the phonon mean free path reduces to half of its wavelength.\cite{beekman2017inorganic} This model assumes that the heat is transported through a solid via random walks between localized osciallators. Cahill \textit{et al.}'s minimum limit model is also known as the amorphous limit model and can be expressed as the following\cite{cahill1992lower}

\begin{equation} 
\renewcommand{\theequation}{S9}
\kappa{_g}  = \bigg(\frac{\pi}{6}\bigg)^{1/3} k{_B}n^{2/3}\bigg[{v}_{L}\bigg(\frac{T}{\theta{_L}}\bigg)^2\int_{0}^{(\theta{_L}/T)}\frac{x^3e^x}{(e^x - 1)^2} dx + 2{v}_{T}\bigg(\frac{T}{\theta{_T}}\bigg)^2\int_{0}^{(\theta{_T}/T)}\frac{x^3e^x}{(e^x - 1)^2} dx\bigg]
\label{eq1}
\end{equation}

\begin{equation} 
\renewcommand{\theequation}{S10}
\theta{_L} = v_{L} \bigg(\frac{h}{2\pi k{_B}}\bigg)(6\pi^2n)^{1/3}
\label{eq1}
\end{equation}

\begin{equation} 
\renewcommand{\theequation}{S11}
\theta{_T} = v_{T} \bigg(\frac{h}{2\pi k{_B}}\bigg)(6\pi^2n)^{1/3}
\label{eq1}
\end{equation}

\noindent where $\kappa_{g}$, \textit{k$_B$}, \textit{n}, \textit{h}, $\theta$, and \textit{T} represent thermal conductivity prediction, Boltzmann constant, number density, Planck constant, Debye temperature, and temperature, respectively. 

Cahill \textit{et al.}'s model has successfully predicted the thermal conductivity of many disordered solids and amorphous materials. However, several recent studies have showed that the thermal conductivity of a few materials can fall below this limit.\cite{chiritescu2007ultralow,duda2013exceptionally,wang2013ultralow,aryana2021tuning} Therefore, to better estimate the theoretical lower bound of thermal conductivity, Agne \textit{et al.}\cite{agne2018minimum} predicted a model based on the diffuson-mediated thermal transport. Agne \textit{et al.}'s diffuson model assumes that the heat is transported through a material via diffusons (i.e., non-propagating, delocalized vibrational modes). According to this diffuson model, the thermal conductivity ($\kappa_{d}$) can be expressed as 

\begin{equation} 
\renewcommand{\theequation}{S12}
\kappa{_d}  = \frac{n^{-2/3}k{_B}}{2\pi^3v_{s}^3}\bigg(\frac{2\pi k{_B}T}{h}\bigg)^4\int_{0}^{0.95\frac{\theta{_D}}{T}}\frac{x^5e^x}{(e^x - 1)^2} dx
\label{eq1}
\end{equation}
\begin{equation} 
\renewcommand{\theequation}{S13}
\theta{_D} = \frac{h}{2\pi k{_B}}(6\pi^2n)^{1/3}v_{s}
\end{equation}
\vspace{0.1mm}

The minimum limit and diffuson limit shown in Figure 2(a) correspond to BaZrS$_{3}$ and are calculated using the properties mentioned in supporting information section S11.

\newpage
\textbf{{{\Large S14. Anisotropic minimum thermal conductivity limit}}}

As RP structured crystals has certain anisotropy, we use the anisotropic thermal conductivity limit proposed by Chen and Dames\cite{chen2015anisotropic} for Ba$_3$Zr$_2$S$_7$. According to this model, the minimum thermal conductivity along the cross-plane direction ($\kappa_{min-c, layered}$ ) of a layered material can be expressed as:  

\begin{equation} 
\renewcommand{\theequation}{S14}
\kappa{_{min-c, layered}}  = \sum_{pol}^{}\frac{k{_B}^{3}}{6\pi({\frac{h}{2\pi})}^2}\frac{{v}_{c}}{{v}_{ab}^2} \bigg[\int_{0}^{\frac{\theta_{D,c}}{T}}\frac{T^2x^3e^x}{(e^x - 1)^2} dx + \frac{\theta_{D,c}^3}{T}\int_{\frac{\theta_{D,c}}{T}}^{\frac{\theta_{D,ab}}{T}}\frac{e^x}{(e^x - 1)^2}\bigg(\frac{\theta_{D,ab}^2 - (Tx)^2}{\theta_{D,ab}^2 - \theta_{D,c}^2}\bigg)^{3/2}dx\bigg]
\label{eq1}
\end{equation}

Where \textit{k$_B$}, \textit{h}, $\theta_{D}$, and \textit{T} represent Boltzmann constant, Planck constant, \textit{c}-axis sound velocity, ab-plane sound velocity, Debye temperature, and temperature, respectively.

For Ba$_3$Zr$_2$S$_7$, we derive \textit{v$_{c,L}$} and \textit{v$_{ab,L}$} (\textit{L} represents longitudinal polarizations) from the elastic constants obtained via MD. The derived values are 4288 and 3807 m s$^{-1}$, respectively. Using these values, the anisotropic minimum thermal conductivity model of Figure S6 is obtained. As elastic constants have an uncertainty of $\sim$12$\%$ and we assume \textit{v$_T$}= 0.6\textit{v$_L$},\cite{aryana2021tuning} the anisotropic model of Figure S6 has an uncertainty of $\sim$15$\%$.  

The anisotropic version of the minimum thermal conductivity model is still higher than the experimental values of Ba$_3$Zr$_2$S$_7$. This can be attributed to the comparable bond strength across the rock-salt layers and (BaZrS$_3$)$_n$ layers. As a result, the sound velocities are also comparable between \textit{c}-axis and \textit{ab}-plane. In other anisotropic materials, such as WSe$_2$, the bonding and sound velocities are very different between \textit{c}-axis and \textit{ab}-plane.\cite{chen2015anisotropic}

Due to comparable sound velocities between \textit{c}-axis and \textit{ab}-plane, the predictions of anisotropic thermal conductivity model are close to the predictions of isotropic thermal conductivity model.  

\begin{figure}[hbt!]
\begin{center}
\renewcommand{\thefigure}{S6}
\includegraphics[scale = 0.28]{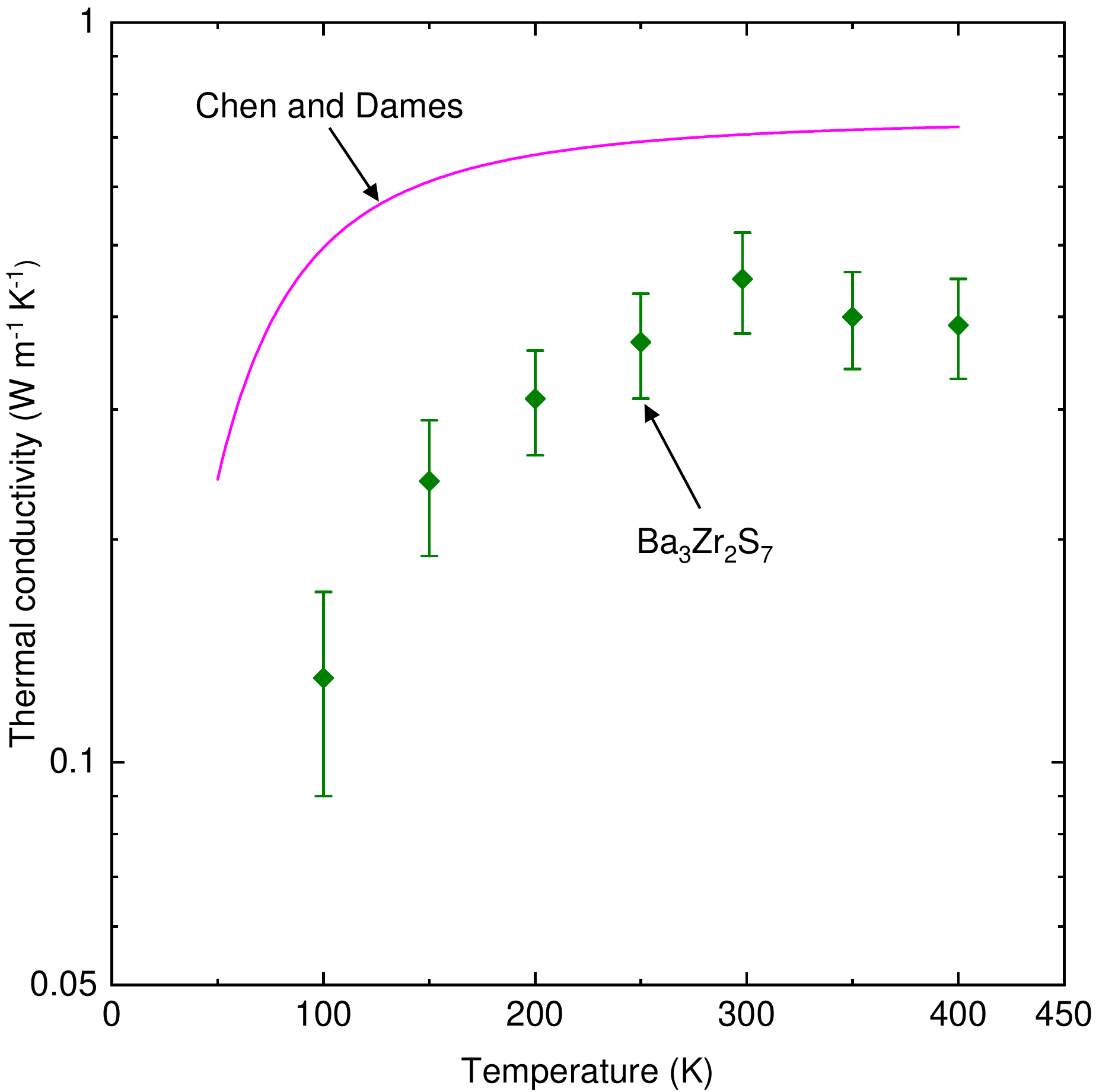}\\
\caption{Anisotropic minimum thermal conductivity model\cite{chen2015anisotropic} and experimental measurements of Ba$_3$Zr$_2$S$_7$.}
\label{Figure 2}
\end{center}
\end{figure}

\newpage
\textbf{{{\Large S15. Machine learning potentials (MLPs)}}}

In order to perform MLMD simulations over a wide temperature range, we develope two machine learning potentials (MLPs) for each of BaZrS$_{3}$ and Ba$_3$Zr$_2$S$_{7}$, respectively. One potential is for relatively low-temperature (LT) MLMDs and the other is for relatively high-temperature (HT) MLMDs. In particular, the LT-MLP of BaZrS$_{3}$ is used to run 100 K, 150 K, 200 K, and 250 K MLMDs, while the HT-MLP of BaZrS$_{3}$ is used to run 300 K, 350 K, and 400 K MLMDs. The LT-MLP of Ba$_3$Zr$_2$S$_{7}$ is for 100 K and 150 K MLMDs, while HT-MLP of Ba$_3$Zr$_2$S$_{7}$ is for 200 K, 250 K, 300 K, 350 K, and 400 K MLMDs.

\vspace{5 mm}
\begin{table*}[hbt!]
\renewcommand{\thetable}{S2}
\large
\centering
\caption{The root-mean-square (RMS) errors of energies per atom and atomic forces for MLPs across both the training and testing datasets.}
\vspace{2 mm}
\setlength{\tabcolsep}{30pt}
\begin{tabular}{ccccc}
\hline
\hline
\multirow{2}{*}{RMS absolute difference} & \multicolumn{2}{c}{MLP (BaZrS$_{3}$)} & \multicolumn{2}{c}{MLP (Ba$_3$Zr$_2$S$_{7}$)}\\
\cline{2-5}
& Train & Test & Train & Test\\
\hline
Energy per atom (meV) & 0.348 & 0.354 & 0.293 & 0.311\\
\hline
Atomic force (eV/\r{A}) & 0.054 & 0.054 & 0.047 & 0.047\\
\hline
\hline
\end{tabular}
 \end{table*}

\newpage
\begin{figure}[hbt!]
\begin{center}
\renewcommand{\thefigure}{S7}
\includegraphics[width=\textwidth]{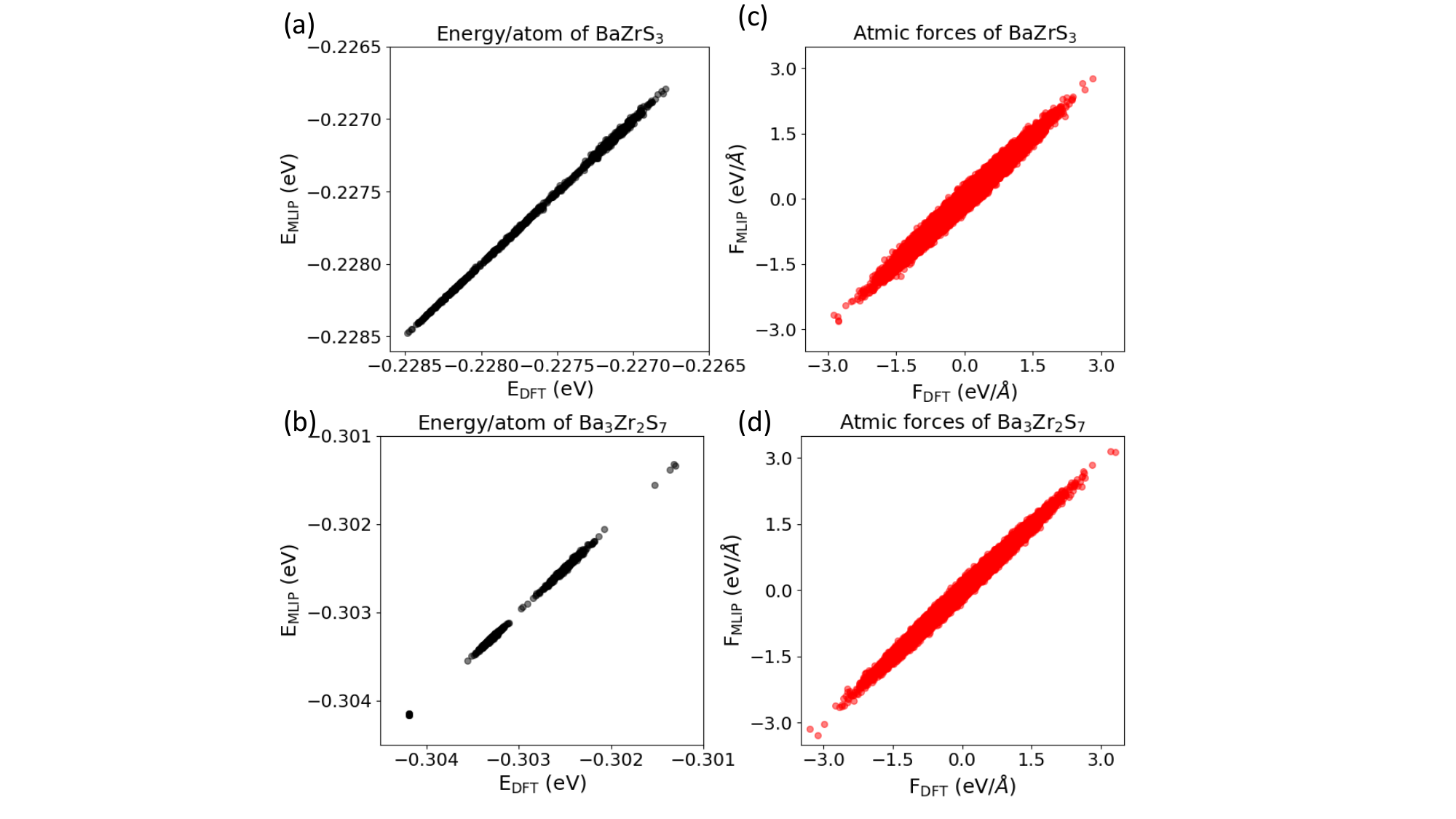}\\
\caption{Discrepancies between the energy/atom and atomic forces obtained using the machine-learned force field (MLIP) and those obtained using DFT over the testing datasets for BaZrS$_{3}$ and Ba$_3$Zr$_2$S$_{7}$.}
\label{Figure 2}
\end{center}
\end{figure}

\newpage
\begin{figure}[hbt!]
\begin{center}
\renewcommand{\thefigure}{S8}
\includegraphics[width=\textwidth]{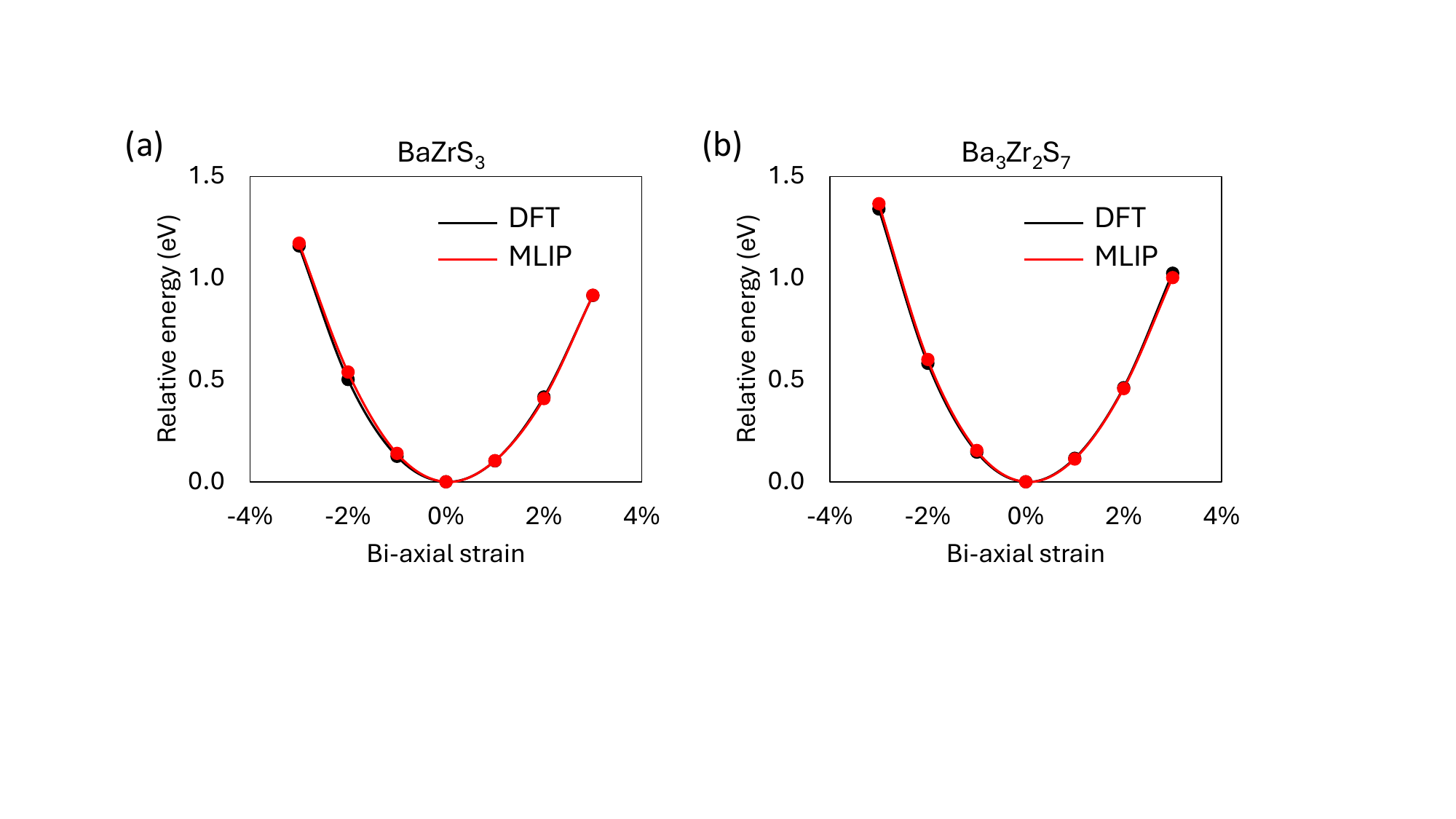}\\
\caption{Calculated relative energies as a function of bi-axial strain between DFT and MLIP for BaZrS$_{3}$ and Ba$_3$Zr$_2$S$_{7}$. The results between DFT and MLIP show very good agreement.}
\label{Figure 2}
\end{center}
\end{figure}

\newpage
\textbf{{{\Large S16. Literature data of elastic modulus/thermal conductivity (E/$\kappa$) ratio}}}

The elastic modulus/thermal conductivity (E/$\kappa$) ratios of a wide range of single crystalline materials are listed in Table S3.\\

\begin{table*}[hbt!]
\renewcommand{\thetable}{S3}
\centering
\caption{The elastic modulus/thermal conductivity (E/$\kappa$) ratios of different groups of single crystalline materials at room temperature.}
\begin{tabular}{cccc}
\hline
\hline\\
Group & Material & E (GPa) & E/$\kappa$ (GPa m K W$^{-1}$)
\vspace{3mm}\\
\hline\\
& Co$_6$S$_8$ & 4 & 18.18\cite{ong2017orientational}
\vspace{1mm}\\
& Co$_6$Se$_8$ & 2.3 & 12.78\cite{ong2017orientational}
\vspace{1mm}\\
Superatom & Co$_6$Te$_8$ & 0.62 & 4.77\cite{ong2017orientational}
\vspace{1mm}\\
& [Co$_6$Se$_8$][C$_{60}$]$_2$ & 8.1 & 32.4\cite{ong2017orientational}
\vspace{1mm}\\
& [Co$_6$Te$_8$][C$_{60}$]$_2$ & 1.5 & 9.38\cite{ong2017orientational}
\vspace{4mm}\\
\hline
\vspace{1mm}\\
& Diamond (I/ IIa/ IIb) & 1144.81 & 1.27/0.49/0.84\cite{mason1972properties,mason1972acoustic,gray1972american,ho1972thermal}
\vspace{1mm}\\
& cBN & 909 & 1.03\cite{grimsditch1994elastic,lehmann2002young,chen2020ultrahigh}
\vspace{1mm}\\
& AlN & 374 & 1.17\cite{yonenaga2002nano,cheng2020experimental}
\vspace{1mm}\\
& GaN & 295 & 1.17\cite{nowak1999elastic,shibata2007high}
\vspace{1mm}\\
& Si & 165.82 & 1.11\cite{mason1972properties,mason1972acoustic,gray1972american,ho1972thermal}
\vspace{1mm}\\
Semiconductor & Ge & 135.4 & 2.25\cite{mason1972properties,mason1972acoustic,gray1972american,ho1972thermal}
\vspace{1mm}\\
& AgSbTe$_2$ & 49.49 & 72.78\cite{morelli2008intrinsically,berri2012ab}
\vspace{1mm}\\
& PbTe & 67.23 & 28.01\cite{miller1981pressure,morelli2008intrinsically}
\vspace{1mm}\\
& InAs & 79.7 & 2.95\cite{gerlich1963elastic,maycock1967thermal}
\vspace{1mm}\\
& PbSe & 65.2 & 24.7\cite{xiao2016origin}
\vspace{1mm}\\
& PbS & 70.2 & 25\cite{xiao2016origin}
\vspace{4mm}\\
\hline\\
& MAPbCl$_3$ (cubic) & 23 & 31.5\cite{elbaz2017phonon}
\vspace{1mm}\\
& MAPbBr$_3$ (cubic) & 17.8 & 34.9\cite{elbaz2017phonon}
\vspace{1mm}\\
Metal halide perovskite & MAPbI$_3$ (tetragonal) & 12 & 35.3\cite{elbaz2017phonon}
\vspace{1mm}\\
& CsPbBr$_3$ (Orthorhombic) & 13.5 & 29.35\cite{elbaz2017phonon}
\vspace{1mm}\\
& FAPbBr$_3$ (Orthorhombic) & 10.2 & 20.82\cite{elbaz2017phonon}
\vspace{4mm}\\
\hline
\hline
\end{tabular}
\end{table*}

\newpage
\begin{table*}[hbt!]
\renewcommand{\thetable}{S3}
\centering
\caption{The elastic modulus/thermal conductivity (E/$\kappa$) ratios of different groups of single crystalline materials at room temperature (cont.).}
\begin{tabular}{cccc}
\hline
\hline\\
Group & Material & E (GPa) & E/$\kappa$ (GPa m K W$^{-1}$)
\vspace{3mm}\\
\hline\\
& MgO & 310 & 5.96\cite{chung1963elastic,braun2018charge}
\vspace{1mm}\\
& Al$_2$O$_3$ & 345 & 10.15\cite{braun2018charge}
\vspace{1mm}\\
& SrTiO$_3$ & 260.85 & 23.71\cite{bell1963elastic,yu2008thermal,oh2011thermal}
\vspace{1mm}\\
& BaZrO$_3$ & 181 & 42.1\cite{vassen2000zirconates}
\vspace{1mm}\\
& La$_2$Zr$_2$O$_7$ & 175 & 92.11\cite{vassen2000zirconates}
\vspace{1mm}\\
& Y$_2$O$_3$-stabilized ZrO$_2$ (YSZ) & 210 & 98.6\cite{vassen2000zirconates}
\vspace{1mm}\\
Oxide & NiO & 175 & 5.15\cite{lewis1973thermal,jifang1991elastic}
\vspace{1mm}\\
& J14 ESO & 152 & 51.53\cite{braun2018charge}
\vspace{1mm}\\
& J34 ESO & 180.8 & 125.56\cite{braun2018charge}
\vspace{1mm}\\
& J36 ESO & 229.9 & 143.69\cite{braun2018charge}
\vspace{4mm}\\
\hline
\vspace{1mm}\\
& Cs$_3$Bi$_2$I$_{6}$Cl$_{3}$ & 17.4 & 87\cite{acharyya2022glassy}
\vspace{1mm}\\
 Layered materials & Cs$_2$PbI$_2$Cl$_{2}$* & 20.6 & 55.7\cite{acharyya2020intrinsically}
\vspace{1mm}\\
& WSe$_2$ & 167.3 & 107.2\cite{chiritescu2007ultralow,zhang2016elastic}

\vspace{4mm}\\
\hline
\vspace{1mm}\\
& BaZrS$_{3}$ & 84.2 & 54.3
\vspace{1mm}\\
& Ba$_3$Zr$_2$S$_{7}$ & 69 & 153.3
\vspace{4mm}\\
\hline
\hline
\end{tabular}
\end{table*}
*Cs$_2$PbI$_2$Cl$_{2}$ is a RP material. It's elastic modulus has been calculated from reported bulk modulus and shear modulus.

\clearpage
\textbf{{{\Large S17. Temperature-dependent thermal conductivity of perovskites}}}

\begin{figure}[hbt!]
\begin{center}
\renewcommand{\thefigure}{S9}
\includegraphics[scale = 0.51]{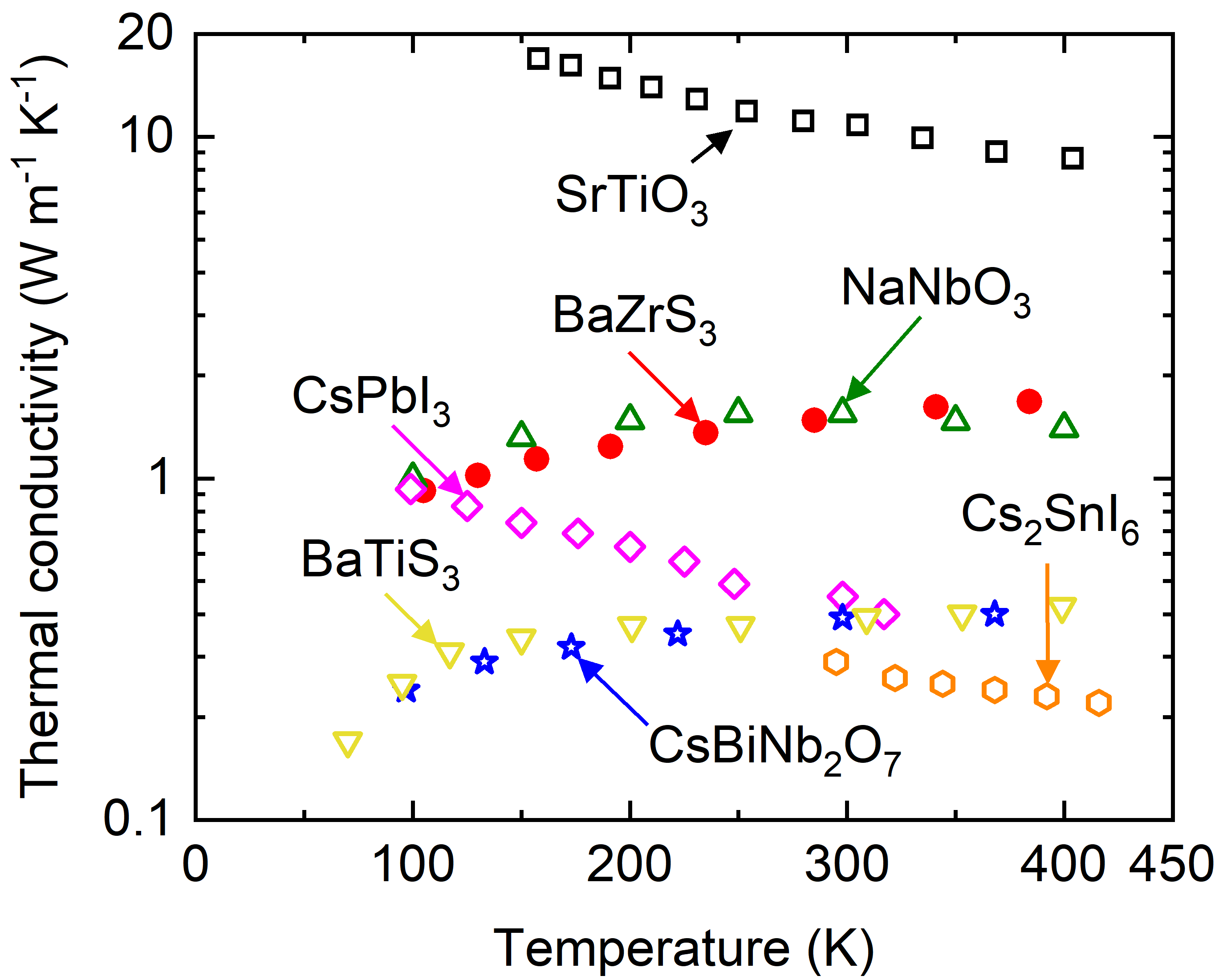}\\
\caption{Temperature-dependent thermal conductivity of BaZrS$_{3}$, SrTiO$_{3}$,\cite{dawley2021thermal} NaNbO$_{3}$,\cite{tachibana2008thermal} CsPbI$_{3}$,\cite{lee2017ultralow} Cs$_{2}$SnI$_{6}$,\cite{bhui2022intrinsically}, BaTiS$_{3}$,\cite{sun2020high} and CsBiNb$_{2}$O$_{7}$.\cite{cahill2010low} The solid and hollow symbols represent measurements taken in this study and literature values, respectively.}
\label{Figure 2}
\end{center}
\end{figure}

\clearpage
\textbf{{{\Large S18. Phonon mean free path of Ba$_{3}$Zr$_{2}$S$_{7}$}}}

\begin{figure}[hbt!]
\begin{center}
\renewcommand{\thefigure}{S10}
\includegraphics[scale = 0.51]{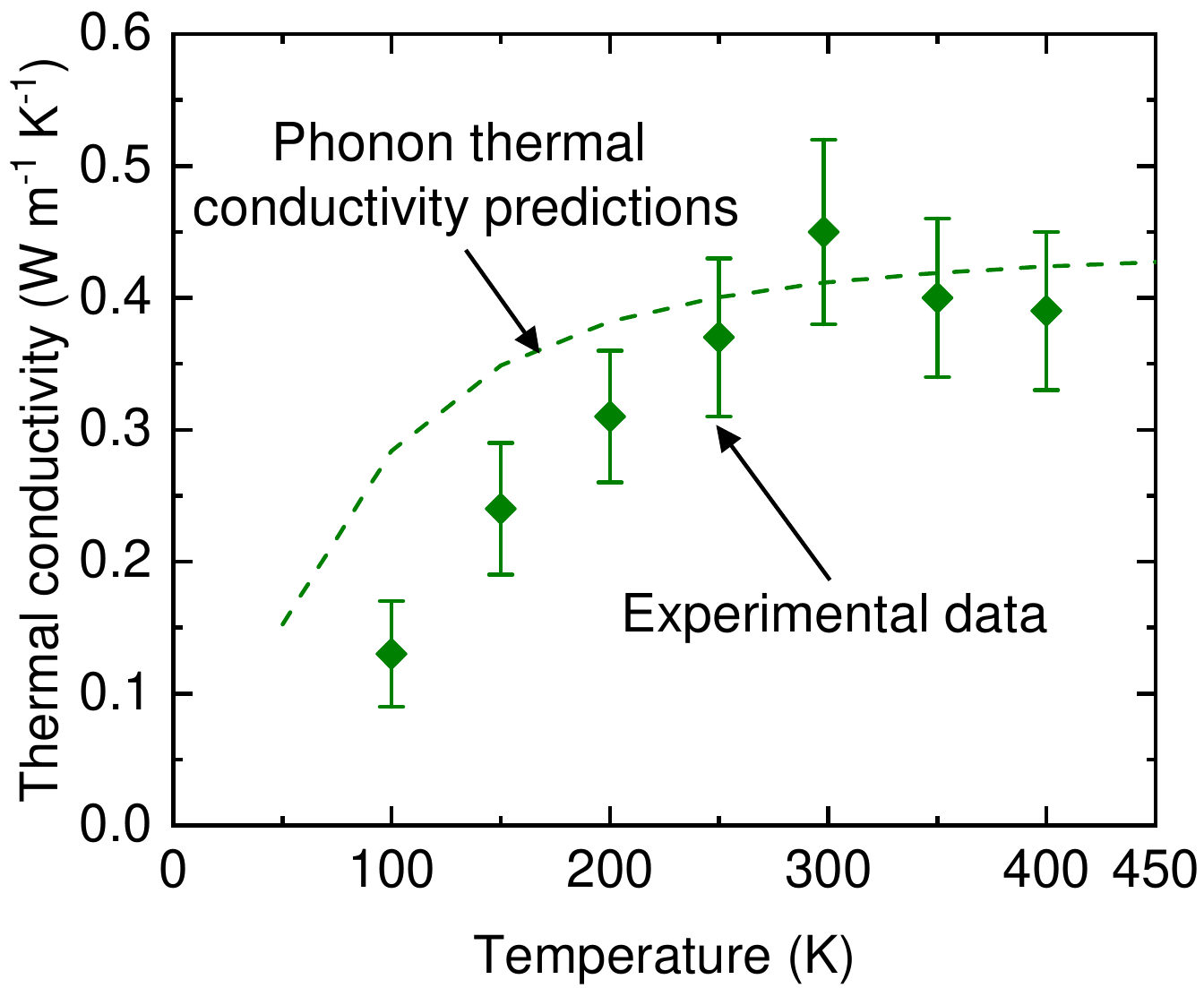}\\
\caption{Comparison of phonon thermal conductivity assuming an average mean free path of 1 nm with experimental data.}
\label{Figure 2}
\end{center}
\end{figure}

\clearpage
\textbf{{{\Large S19. Heat-current autocorrelation functions}}}

\begin{figure}[hbt!]
\begin{center}
\renewcommand{\thefigure}{S11}
\includegraphics[width=\textwidth]{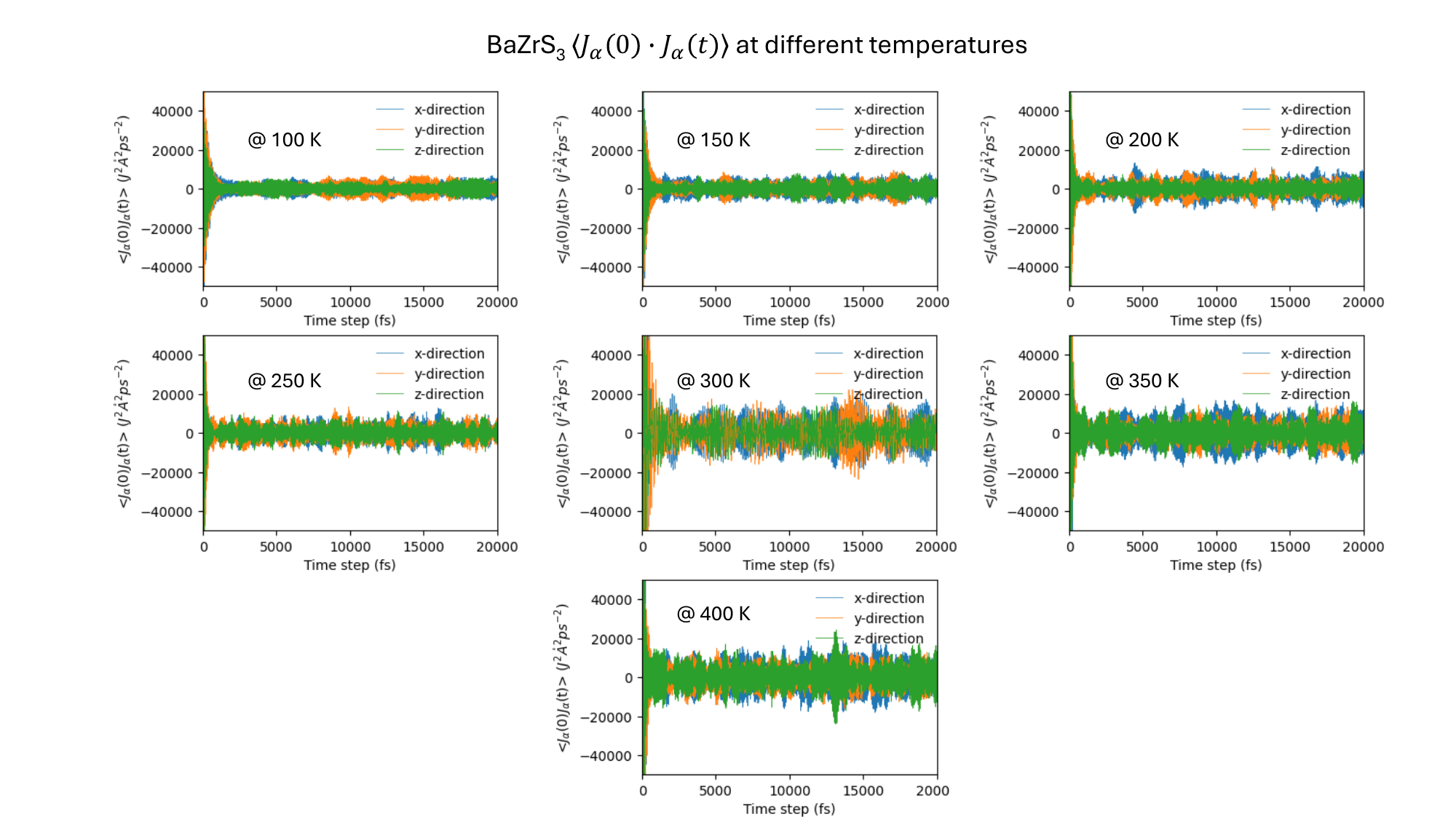}\\
\caption{The calculated heat-current autocorrelation functions for BaZrS$_{3}$ at various considered temperatures. Recall that we performed eight independent MLMDs at each temperature. For simplicity, each panel in the figure represents one MLMD simulation case out of these eight independent cases. Blue, orange, and green colors represent x, y, and z-direction results, respectively.}
\label{Figure 2}
\end{center}
\end{figure}

\clearpage
\begin{figure}[hbt!]
\begin{center}
\renewcommand{\thefigure}{S12}
\includegraphics[width=\textwidth]{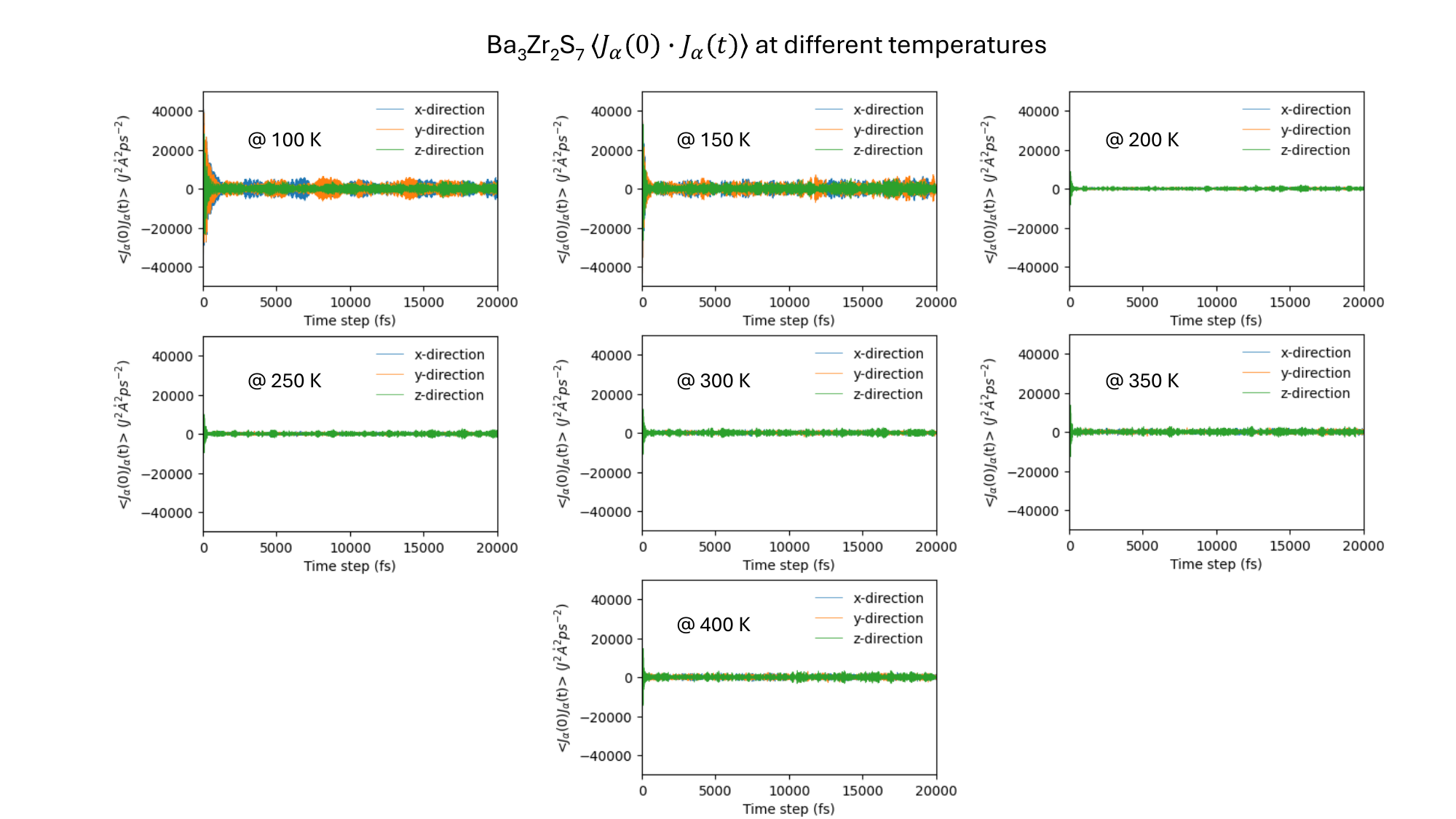}\\
\caption{The calculated heat-current autocorrelation functions for Ba$_{3}$Zr$_{2}$S$_{7}$ at various considered temperatures. Recall that we performed eight independent MLMDs at each temperature. For simplicity, each panel in the figure represents one MLMD simulation case out of these eight independent cases. Blue, orange, and green colors represent x, y, and z-direction results, respectively.}
\label{Figure 2}
\end{center}
\end{figure}

\clearpage
\textbf{{{\Large S20. Calculated thermal conductivities vs correlation time}}}

\begin{figure}[hbt!]
\begin{center}
\renewcommand{\thefigure}{S13}
\includegraphics[width=\textwidth]{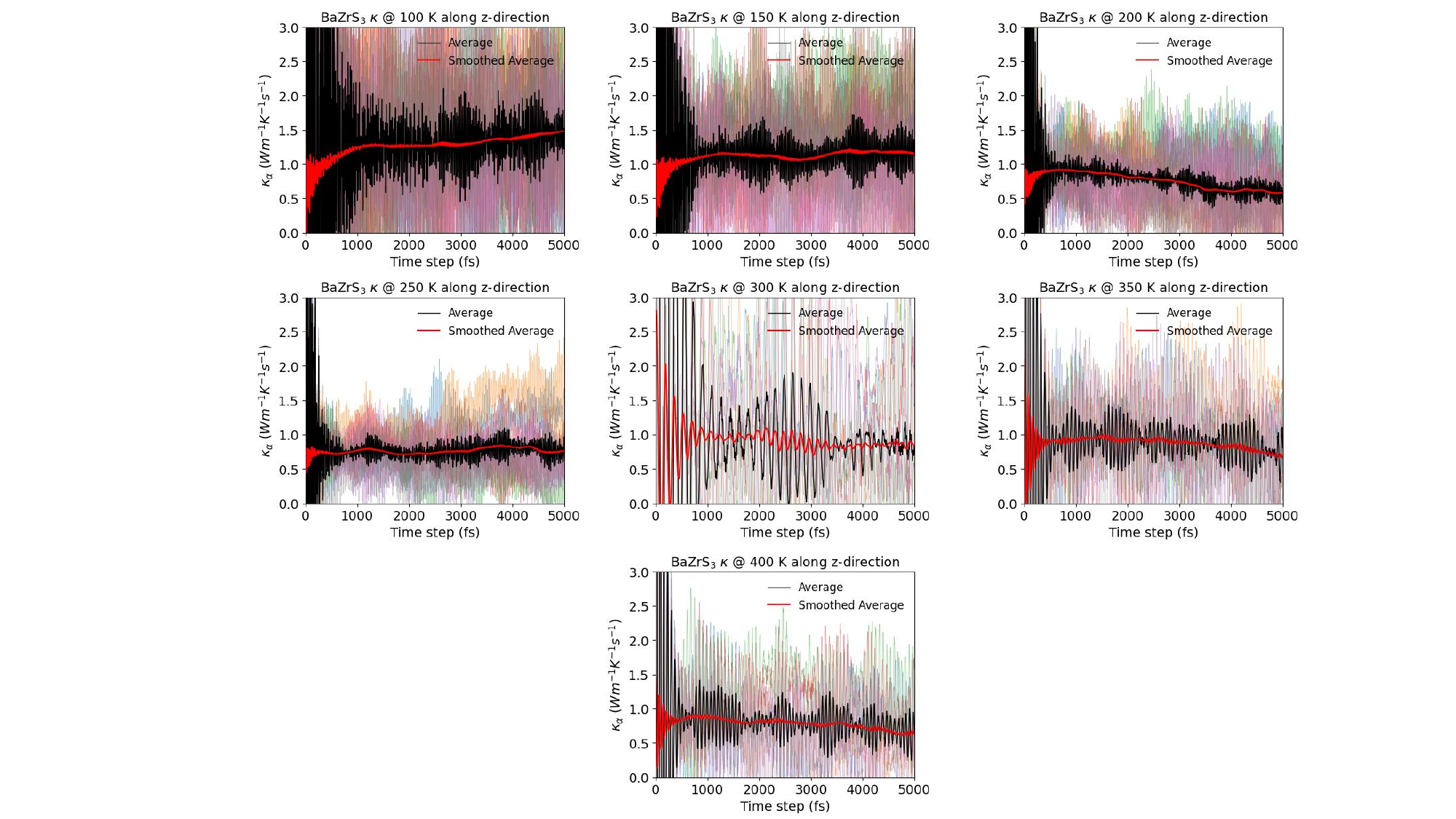}\\
\caption{The calculated thermal conductivities of BaZrS$_{3}$ as a function of correlation time at various considered temperatures. Only the z-components of the thermal conductivity are plotted. Within each panel, the colorful curves represent the calculated results for all eight cases at each temperature. The black curves depict the average of the eight colorful curves, while the red curves are obtained by averaging every 200 data points from the black curve.}
\label{Figure 2}
\end{center}
\end{figure}

\clearpage
\begin{figure}[hbt!]
\begin{center}
\renewcommand{\thefigure}{S14}
\includegraphics[width=\textwidth]{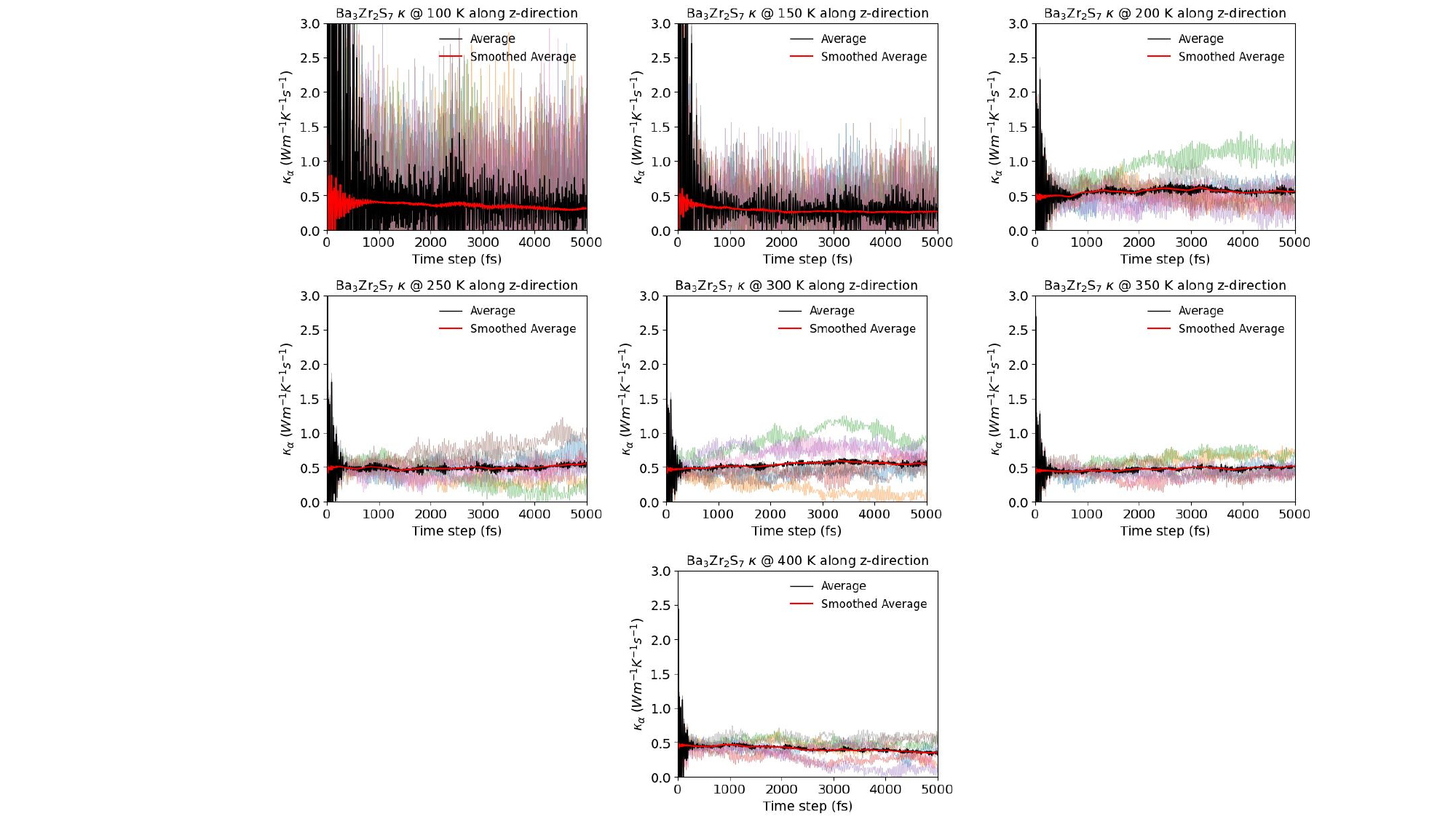}\\
\caption{The calculated thermal conductivities of Ba$_{3}$Zr$_{2}$S$_{7}$ as a function of correlation time at various considered temperatures. Only the z-components of the thermal conductivity are plotted. Within each panel, the colorful curves represent the calculated results for all eight cases at each temperature. The black curves depict the average of the eight colorful curves, while the red curves are obtained by averaging every 200 data points from the black curve.}
\label{Figure 2}
\end{center}
\end{figure}

\clearpage
\textbf{{{\Large S21. DFT-calculated phonon dispersion}}}

\begin{figure}[hbt!]
\begin{center}
\renewcommand{\thefigure}{S15}
\includegraphics[width=\textwidth]{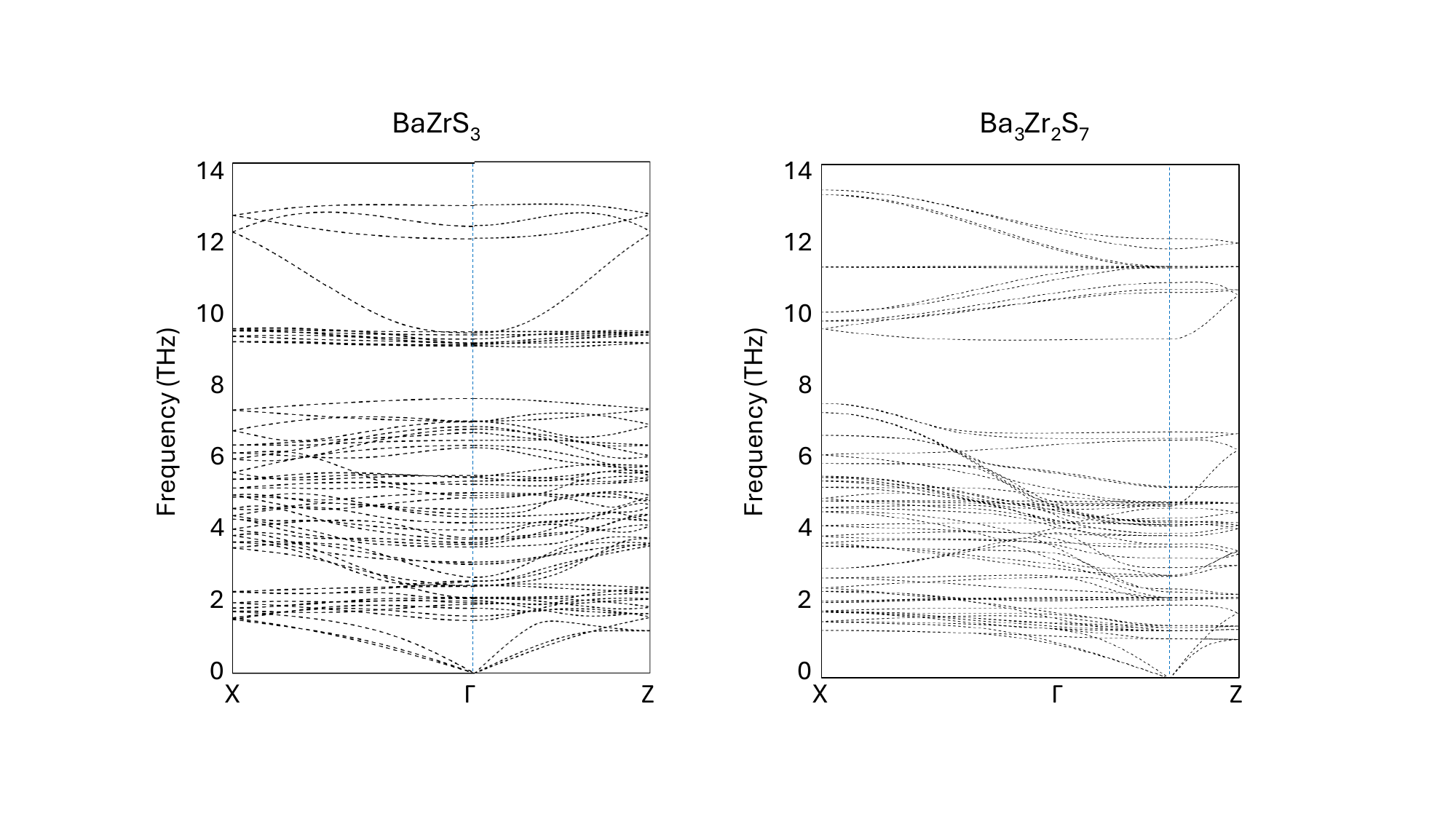}\\
\caption{DFT-calculated phonon dispersion of BaZrS$_{3}$ and Ba$_{3}$Zr$_{2}$S$_{7}$. Note that they are in reasonable agreement with the SED results.}
\label{Figure 2}
\end{center}
\end{figure}

\clearpage
\textbf{{{\Large S22. Eigen-displacements of phonon modes}}}

\begin{figure}[hbt!]
\begin{center}
\renewcommand{\thefigure}{S16}
\includegraphics[width=\textwidth]{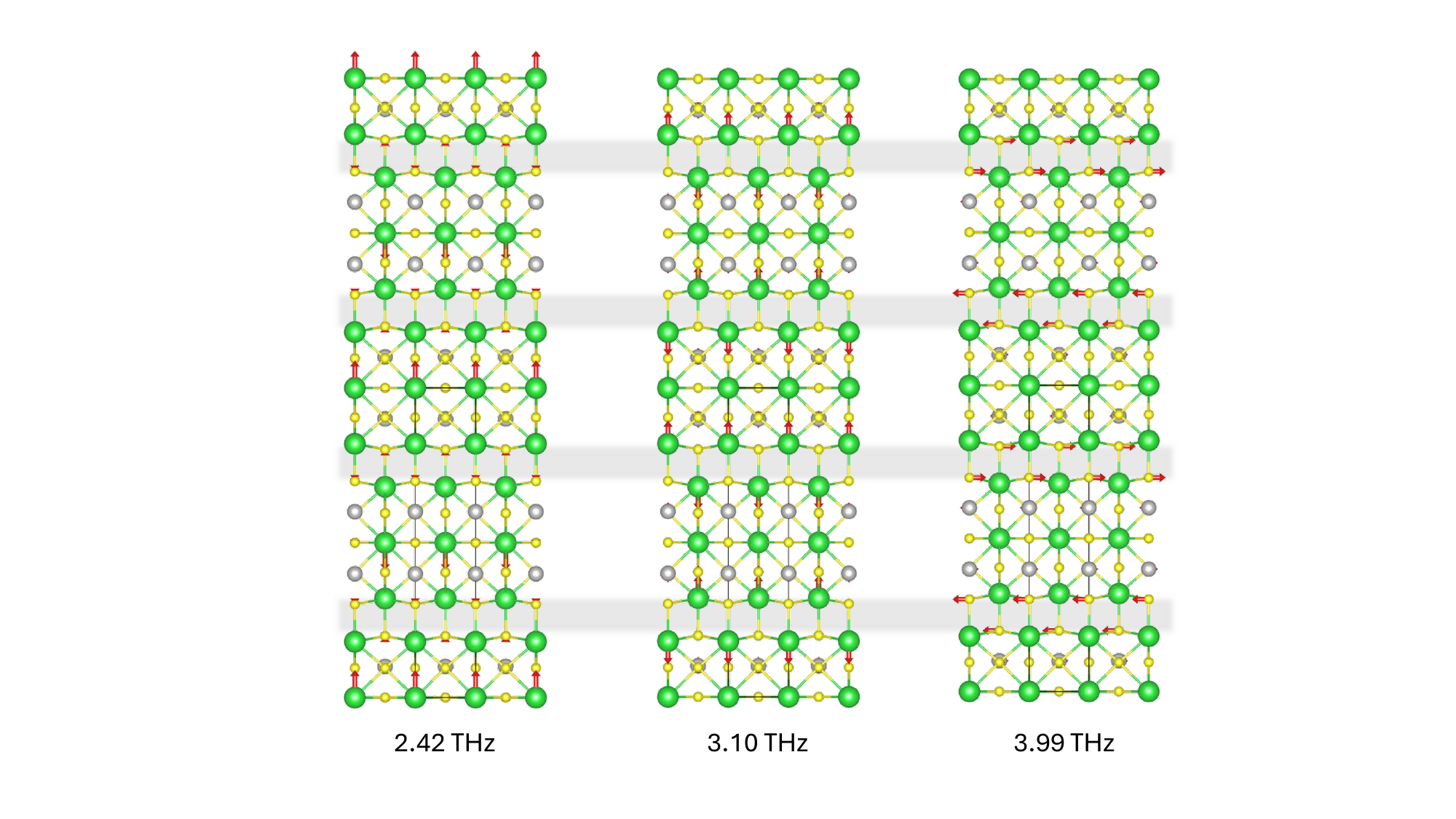}\\
\caption{Eigen-displacements of three prototype phonon modes for Ba$_{3}$Zr$_{2}$S$_{7}$, illustrating that the inter- and intra-cell rock-salt blocks contribute to phonon localizations. The gray bars highlight the rock-salt-like interfaces. }
\label{Figure 2}
\end{center}
\end{figure}

\medskip

%
\newpage
\bibliographystyle{achemso}
\bibliography{BZS}




\end{document}